\documentclass[prd,preprintnumbers,aps,preprint,amsmath,amssymb,superscriptaddress,floatfix,nofootinbib,unsortedaddress]{revtex4}
\pdfoutput=1

\usepackage{amssymb, bm, amsmath, graphicx, physymb}
\usepackage[colorlinks=true,linktocpage=true,linkcolor=blue,citecolor=blue]{hyperref}
\usepackage[usenames,dvipsnames]{color}
\usepackage{slashed}
\usepackage[utf8]{inputenc}
\usepackage[T1]{fontenc}
\usepackage{enumerate}
\usepackage{amsfonts}
\usepackage{mathtools}
\usepackage{latexsym}
\usepackage{hyperref}
\usepackage{epstopdf} 
\usepackage{bbm}
\usepackage{soul}
\usepackage[normalem]{ulem}
\usepackage{braket}
\usepackage{nccmath}
\usepackage{graphicx}

\usepackage[e]{esvect}

\def\cN{{\cal N}}
\def\cR{{\cal R}}
\def\cO{{\cal O}}

\def\cQ{{\cal Q}}

\def\beps{{\boldsymbol \epsilon}}

\newcommand{\secn}[1]{Section~1}
\newcommand{\appn}[1]{Appendix~1}

\long\def\comment#1{ }

\def\and{\quad\text{and}\quad}

\def\g{{\boldsymbol g}}
\def\q{{\boldsymbol q}}
\def\0{{\boldsymbol 0}}
\def\1{{\boldsymbol 1}}
\def\p{{\boldsymbol p}}
\def\l{{\boldsymbol l}}
\def\k{{\boldsymbol k}}
\def\x{{\boldsymbol x}}

\def\u{{\boldsymbol u}}

\def\b{{\boldsymbol b}}
\def\0{{\boldsymbol 0}}
\def\Q{{\boldsymbol Q}}

\renewcommand\a{\alpha}
\renewcommand\b{\beta}
\renewcommand\d{\delta}
\renewcommand\l{\lambda}

\renewcommand\u{\upsilon}

\renewcommand\o{\omega}

\newcommand\m{\mu}

\newcommand{\re}{{\rm{Re}}}
\newcommand{\im}{{\rm{Im}}}

\def\u{{\boldsymbol u}}
\newcommand{\tvec}{\boldsymbol}

\renewcommand{\part}{{\rm part}}

\newcommand{\be}{\begin{equation}}
\newcommand{\ee}{\end{equation}}
\newcommand{\bes}{\begin{subequations}}
\newcommand{\ees}{\end{subequations}}
\newcommand{\bea}{\begin{eqnarray}}
\newcommand{\eea}{\end{eqnarray}}

\newcommand{\pa}{\partial}

\newcommand{\na}{\nabla}

\begin{document}

\title{Jet quenching in anisotropic flowing matter}

\author{Matvey V. Kuzmin}
\email[Email: ]{kuzmin.mv19@physics.msu.ru}
\affiliation {NRC Kurchatov Institute, Moscow, Russia}
\affiliation {Faculty of Physics, Moscow State University, Moscow 119991, Russia}
\author{Xo{\'{a}}n Mayo L\'{o}pez}
\email[Email: ]{xoan.mayo.lopez@usc.es}
\affiliation{Instituto Galego de F{\'{i}}sica de Altas Enerx{\'{i}}as,  Universidade de Santiago de Compostela, Santiago de Compostela 15782, Galicia, Spain}
\author{Jared Reiten}
\email[Email: ]{jdreiten@physics.ucla.edu}
\affiliation{Department of Physics and Astronomy, University of California, Los Angeles, CA 90095, USA}
\affiliation{Mani L. Bhaumik Institute for Theoretical Physics, University of California, Los Angeles, CA 90095, USA}
\author{Andrey V. Sadofyev}
\email[Email: ]{andrey.sadofyev@usc.es}
\affiliation{Instituto Galego de F{\'{i}}sica de Altas Enerx{\'{i}}as,  Universidade de Santiago de Compostela, Santiago de Compostela 15782, Galicia, Spain}
\affiliation{LIP, Av. Prof. Gama Pinto, 2, P-1649-003 Lisboa, Portugal}

\begin{abstract}
We study the interplay between the flow and hydrodynamic gradients in jet quenching at first order in opacity. We find that the mixed flow-gradient contributions in jet quenching are
enhanced by the medium length, and survive in the eikonal limit, dominating over other medium evolution effects. The resulting modification to the jet quenching parameter and energy loss rate can be substantial, leading to ample phenomenological implications. We also compute the leading corrections to the jet broadening due to the flow velocity gradients, and consider the leading gradient effects in the medium-induced branching for general kinematics, extending the recent considerations of jets in inhomogeneous media. These results can be straightforwardly coupled to matter simulations, providing new opportunities for jet tomography in heavy-ion collisions.
\end{abstract}

\maketitle

\section{Introduction}

In-medium energy loss and the related suppression of energetic partons provide primary evidence for the formation of collective matter in heavy-ion collisions (HIC), for a review see e.g. \cite{Busza:2018rrf,Cunqueiro:2021wls, Apolinario:2022vzg}. Jets formed through the branching of such energetic partons interact with the resulting matter across multiple length scales, and are sensitive to the details of the medium evolution. Thus, the medium-induced modification of jets can be used for imaging of the quark-gluon plasma (QGP) created in HIC, laying the foundation of so-called jet tomography, see e.g. \cite{Vitev:2002pf, Wang:2002ri, JET:2013cls, Betz:2014cza, Xu:2014ica, Djordjevic:2016vfo, Apolinario:2017sob, Apolinario:2020uvt, Du:2021pqa} and references therein.

Describing the interaction of energetic partons with the medium within the framework of perturbative QCD, one typically characterizes the matter with a collective background color field, see e.g. \cite{Gyulassy:1993hr, Zakharov:1996fv, Baier:1996kr,  Wiedemann:2000ez, Wiedemann:2000za, Gyulassy:2000er, Gyulassy:2002yv, Arnold:2002ja}. Partons are then deflected by the matter field and lose energy in the process, mainly through gluon bremsstrahlung. The matter field is inherently stochastic, and hence all observables should be averaged over its possible configurations. The resulting formalisms are usually treated under multiple simplifying assumptions, such as the eikonal approximation (the limit of infinitely energetic leading parton) or the limit of static, transversely\footnote{The longitudinal/transverse directions are defined with respect to the momentum of the leading parton.} homogeneous matter. Consequently, the description of jet quenching decouples from the evolution and structure of the matter (especially in the transverse directions), see the discussion in \cite{Sadofyev:2021ohn}. 

There have been multiple attempts to extend the theoretical approaches to jet quenching by including medium-evolution effects with minimal modifications: from taking matter dilution into account \cite{Gyulassy:2000gk, Gyulassy:2001kr,Baier:1998yf}, or using basic kinematic arguments \cite{Baier:2006pt, Liu:2006he, Renk:2006sx}, to treating the transverse flow within phenomenolgically-motivated models \cite{Armesto:2004pt, Armesto:2004vz}. More recently, the theory of jet-matter interactions has been extended to the case of an evolving medium, including the effects of flow \cite{Sadofyev:2021ohn,Sadofyev:2022hhw,Antiporda:2021hpk,Andres:2022ndd} and anisotropies \cite{He:2020iow, Sadofyev:2021ohn,Ipp:2020mjc, Hauksson:2021okc,Carrington:2021dvw,Sadofyev:2022hhw,Barata:2022krd,Fu:2022idl,Barata:2022utc,Carrington:2022bnv,Hauksson:2023tze,Boguslavski:2023alu,Barata:2023qds,Barata:2023zqg}. 
In these works, flow- and anisotropy-induced effects are considered separately. However, such considerations already allow one to probe the coupling of jets with a variety of features of the medium evolution. 

In this work, we focus on the interplay between the effects of flow and anisotropy in jet quenching calculations. Following \cite{Sadofyev:2021ohn}, we work within the opacity expansion, and consider jet momentum broadening and medium-induced gluon radiation in an anisotropic flowing matter. First, we extend the formalism developed in \cite{Sadofyev:2021ohn}, where the structure of matter is treated within a gradient expansion\footnote{See also \cite{Lekaveckas:2013lha, Rajagopal:2015roa, Sadofyev:2015hxa, Reiten:2019fta, Arefeva:2020jvo} for other applications of the gradient expansion to probe-matter interactions.}. In \cite{Sadofyev:2021ohn} the gradient effects are considered only for momentum broadening and in the absence of flow or flow velocity gradients. Here, we start by including the effects of all hydrodynamic gradients at first order in the gradient expansion into the jet momentum broadening consideration, going beyond the discussion in \cite{Sadofyev:2021ohn,Barata:2022krd}. Next, we derive the leading gradient corrections to medium-induced gluon radiation for general kinematics, extending the results available in the literature, {\it c.f.} \cite{Sadofyev:2021ohn, Barata:2023qds}. Finally, we show that the mixed flow-gradient contributions modify the final-state parton distributions already in the eikonal limit, and, moreover, these terms are additionally enhanced by the medium length. Thus, they dominate over other medium evolution effects. As a simple illustration of the effect of these mixed terms, we show that the jet quenching parameter in flowing anisotropic matter reads
\begin{align}
\hat{q}(z)=\left[1 - z\,\hat{\g} \cdot \frac{\u}{1-u_z} \right]\hat{q}_0(z)\,, \notag
\end{align}
where $z$ is the path length of the energetic parton, the transverse gradients are compactly introduced with $\hat{\g}_\a \equiv \left(\boldsymbol{\na}_\a T \frac{\delta }{\delta T} + \boldsymbol{\na}_\a u_z \frac{\delta }{\delta u_z} + \boldsymbol{\na}_\a\u_\b \frac{\delta }{\delta \u_\b}  \right)$, where $T$ is the temperature, $\u$ and $u_z$ are the transverse and longitudinal velocity components, $\hat{q}_0$ is the jet quenching parameter of a static homogeneous matter with the same local properties, and all these objects should be understood as functions of $z$. We argue that such modifications in jet quenching calculations can be substantial for characteristic evolution of the QGP in HIC.

\section{Momentum broadening}

In this work, we study the broadening and medium-induced radiation pattern of a highly-energetic parton interacting with nuclear matter within opacity expansion, closely following the formalism developed in \cite{Sadofyev:2021ohn,Barata:2022krd,Andres:2022ndd,Barata:2022utc,Barata:2023qds}. Before turning to particular processes, let us first specify the details of the setup used in this work. Here, we will ignore the energy-suppressed spin effects, and work with scalar particles in the fundamental representation. In contrast to \cite{Sadofyev:2021ohn}, where the medium-induced emission in flowing matter is studied in the case of a scalar ``gluon,'' we focus on the emission of actual spin-1 gluons of the underlying gauge theory --- see e.g. \cite{Barata:2023qds} for a discussion. 

The matter is modeled by a classical stochastic color field generated by moving massive quasi-particle sources, neglecting their recoil. This background field can be written as
\begin{equation}
    \label{Amu}
    gA^{a\mu}_{\text{ext}}(q) =  \sum_i \, u_i^\mu \, e^{-i(\q\cdot\x_i + q_z z_i)} \, t^a_i \, v_i(q) \; (2\pi)\, \delta(q_0-\q\cdot\u_i-q_z u_{zi})\, ,
\end{equation}
where $u_\mu=(1,\, \u,\, u_z)$ is the non-relativistic velocity (i.e., it is the four-velocity with the relativistic $\gamma$-factor removed), $v_i(q)$ is the single-source potential, $t^a_i$ controls the color of the given source, $(\x_i, z_i)$ are the spatial coordinates of the $i$th source, and the sum runs over all the sources in the medium, while the matter properties can change from point to point. We will use compact notation --- suppressing the $i$ subscript --- where doing so leads to no confusion. 

The particular scattering potential $v(q)$ is model-dependent, and there exist multiple choices in the literature, see e.g. the discussion in \cite{Sadofyev:2021ohn,Antiporda:2021hpk} Throughout this paper, we will use the  Gyulassy-Wang (GW) model to make results more explicit, although generalization is straightforward. The corresponding potential reads
\begin{equation}
\label{v for the GW model}
    v_i(q)=\frac{g^2}{q^2 -\mu_i^2}\, ,
\end{equation}
where $g$ is the effective strong coupling inside the medium, and $\mu_i$ is the Debye mass at the position of the given source. 

Finally, one must specify how the final distributions are averaged over configurations of the background field, fixing the structure of multi-point stochastic correlators. Following \cite{Sadofyev:2021ohn}, we assume the color fields to have Gaussian statistics, taking only pairwise averages into account and enforcing color neutrality, as it is often used in perturbative QCD calculations. The averaging over color sources then results in
\begin{equation}
\langle t^a_i t^b_j\rangle=\frac{1}{2C_{\bar R}}\d_{ij}\d^{ab}\,,
\end{equation}
where $C_{\bar R}$ is the quadratic Casimir in the representation opposite to the representation of the color sources. Here, we will assume that all the sources are in the same fundamental representation, setting $C_{\bar R}=N_c$.

\begin{figure}
    \centering
	\includegraphics[height=3.5cm]{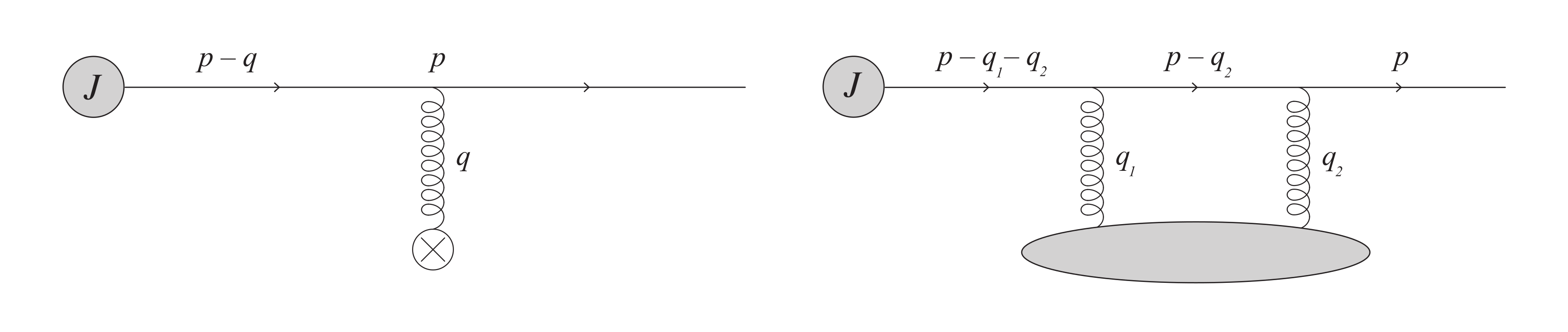}
	\vspace*{-10mm}\caption{The single-Born amplitude $M_1$ (left) and the double-Born amplitude $M_2$ (right) for transverse-momentum broadening.}
	\label{f:Broad}
\end{figure}

In this section, we further derive the gradient correction to the momentum-broadening distribution within flowing, anisotropic QCD matter to first order in opacity, going beyond the discussion in \cite{Sadofyev:2021ohn}. We focus particularly on the interplay between velocity corrections and spatial gradients in the matter variables at first subleading order, neglecting anything suppressed by two or more powers of the initial energy. At first order in opacity $N=1$, up to two interactions in the amplitude squared, there are two types of contributions to take into account: the single-Born (SB) contribution $M_1$ and the double-Born (DB) contribution $M_2$, see Fig.~\ref{f:Broad}. Using this decomposition of the amplitude and omitting the terms whose average vanishes, we can write its square as
\begin{equation}
    \langle|M_{N=1}|^2\rangle = \langle|M_1|^2\rangle + \langle M_2 M_0^*\rangle + \langle M^*_2 M_0\rangle \, ,
\end{equation}
where $M_0$ corresponds to vacuum propagation.

%
%

\subsection{Single-Born contribution}
We start by considering the SB contribution to the broadening of an energetic (scalar) quark. The corresponding $N=1$ amplitude reads 
\begin{align}
\label{M^2atN1}
    iM_1 &= \int \frac{d^4q}{(2\pi)^4} \left[it^a_{proj} (2p-q)_\mu A^{a\mu}_{\text{ext}}(q)\right] \left[\frac{i}{(p-q)^2+i\epsilon}\right] J(p-q) \notag
    \\&= -\sum_i\,t^a_{proj} t^a_i\,\left(1-\frac{\u\cdot\p}{(1-u_z)E}\right)\, \int_q e^{-i(\q \cdot \x_i + q_z z_i)}\, v(q)\, \frac{2(1-u_z)E}{(p-q)^2+i\epsilon} \,J(p-q)\, ,
\end{align}
where we have used the delta function in \eqref{Amu}. In the above equation, $J(p-q)$ is the source of the initial energetic quark controlling the initial distribution of energetic partons --- assumed to be centered at $\x_0=0$ and $z_0=0$ --- and $t^a_{proj}$ is the leading parton (``projectile'') color generator. We have also introduced shorthand notation for integrals running over the full three-dimensional space as $\int_x\equiv d^3 x$ (and $\int_k \equiv \frac{d^3 k}{(2\pi)^3}$), and over the transverse space as $\int_{\x} \equiv \int d^2\x$ (and $\int_{\k} \equiv \int \frac{d^2 \k }{(2\pi)^2}$). 

Integrating over $q_z$ by residues, we assume that $J$ is slowly varying, hence there are only four poles in \eqref{M^2atN1} to account for in the evaluation of the integral. The two poles, coming from the scattering potential $v(q)$, have a finite imaginary part. For a sufficiently dilute and longitudinally-extended medium $\mu_i z_i\gg1$, and so the corresponding contributions are exponentially suppressed and can be neglected. The other two poles come from the quark propagator
\begin{subequations}
\begin{align}
     Q_{p-q}^+ &\simeq \frac{2 E}{1+u_z} \left(1- \frac{\q\cdot\u}{2E}\right) \label{QMpq}\,  , 
     \\  Q_{p-q}^- &\simeq \frac{\q\cdot\u}{1-u_z} + \frac{(\p-\q)^2-\p^2}{2(1-u_z)E} \label{Qmpq} \, ,
\end{align}
\end{subequations}
where we have accounted for the first terms in the eikonal expansion. The pole $Q^+_{p-q}$ is $\mathcal{O}\left(E\right)$, so its residue will be highly suppressed by the leading parton energy and can be neglected. Therefore, the only non-vanishing contribution comes from the residue corresponding to the pole at $Q^-_{p-q}$. The integration contour has to be closed below the real axis enforcing $z_i>0$. After performing the $q_z$-integration, the SB contribution to the amplitude reduces to
\begin{align}
    iM_1 &= -i\sum_i\,t^a_{proj} t^a_i\,\left(1-\frac{\u\cdot\p}{(1-u_z)E}\right)\, \notag
    \\ &\hspace{2cm}\times \int_{\q} \, \theta(z_i)\, e^{-i\q\cdot\x_i}\, e^{-iQ^-_{p-q}z_i}\, v(\tilde{q})\,\frac{2E}{1+u_z}\, \frac{1}{Q^-_{p-q}-Q^+_{p-q}} \,J(p-\tilde{q})\, \\ \notag
    &= i\sum_i\,t^a_{proj} t^a_i\, 
    \int_{\q} \,\theta(z_i)\, e^{-i\q\cdot\x_i}\, e^{-iQ^-_{p-q}z_i}\, \left(1-\frac{\u\cdot(\p-\q)}{(1-u_z)E}\right) v(\tilde{q})\, \,J(p-\tilde{q})\, , 
\end{align}
where the tilde indicates that $q_0$ and $q_z$ have been fixed, and $\tilde{q}_{\mu}=\left(\q\cdot\u+Q^-_{p-q} u_z, \,\q, \,Q^-_{p-q}\right)_\mu$. 

We can now square the amplitude, average over initial and sum over final quantum numbers, as well as average over the field configurations. We will also assume that the sources are smoothly distributed, replacing any sum over the scattering centers by a continuous integral
\begin{align}
    \sum_i f_i &=
    \int_{\x,z} \: \rho(\x,z) \: f(\x,z) \, ,
\end{align}
where $\rho(\x, z)$ is the source number density. Doing so yields
 \begin{align}
 \label{M1^2W/OGamma}
     \langle|M_1|^2\rangle &= \mathcal{C}\,\int_{x,\q,\q'} \, \theta(z) \, \rho(\x,z) \,e^{-i(\q-\q')\cdot\x}\, e^{-i\left(Q^-_{p-q}-Q^-_{p-q'}\right)z}\notag
     \\&\hspace{0.5cm}\times\left(1-\frac{\u\cdot(\p-\q)}{(1-u_z)E}-\frac{\u\cdot(\p-\q')}{(1-u_z)E}\right) \, v(\tilde{q}) \, v^*(\tilde{q}') \, J(p-\tilde{q}) \, J^*(p-\tilde{q}') \, ,
 \end{align}
where $\mathcal{C}=\frac{C_F}{2N_c}$. Note that we only explicitly show the arguments of $\rho(\x,z)$ for illustrative purposes, keeping the arguments of the other hydrodynamic variables implicit.

Contrary to the case with vanishing medium velocity, the $q_z$ integration results in a shift in the argument of the potential and initial source. Expanding these functions, one readily finds
\begin{subequations}
\begin{align}
\label{vArgShift}
    &v(\tilde{q}) \simeq v(\q^2)\left(1+\frac{\q\cdot\u}{(1-u_z)E}\frac{(\p-\q)^2-\p^2}{v(\q^2)} \frac{\partial v}{\partial\q^2}\right)\, ,\\
\label{JArgShift}
    &J(p-\tilde{q}) \simeq J\left(E,\p-\q\right)\left(1- \frac{\q\cdot\u}{(1-u_z)}\frac{1}{J}\frac{\partial J}{\partial E}\right) \, ,
\end{align}
\label{ArgShift}
\end{subequations}
where $v(\q^2)=-\frac{g^2}{\q^2+\m^2}$.

Upon inspection of \eqref{M1^2W/OGamma} and \eqref{ArgShift}, one sees that all the velocity corrections appearing in the squared amplitude are proportional to the transverse velocity $\u$. Hence, it is useful to define a new vector in the transverse plane 
\begin{align}
    \boldsymbol{\Gamma}(\q,\q')\equiv &-\frac{\p-\q}{(1-u_z)E}-\frac{\p-\q'}{(1-u_z)E} +\frac{\q}{(1-u_z)E}\frac{(\p-\q)^2-\p^2}{v(\q^2)} \frac{\partial v}{\partial\q^2} \notag 
    \\ &+\frac{\q'}{(1-u_z)E}\frac{(\p-\q')^2-\p^2}{v^*(\q'^2)} \frac{\partial v^*}{\partial\q'^2} 
    - \frac{\q}{(1-u_z)} \frac{1}{J}\frac{\partial J}{\partial E}- \frac{\q'}{(1-u_z) }\frac{1}{J}\frac{\partial J}{\partial E}\, ,
\end{align}
which enables us to write the averaged squared-amplitude in a compact form
 \begin{align}
 \label{M_1^2 before expanding}
     \langle|M_1|^2\rangle &= \mathcal{C}\,\int_{x,\q,\q'} \, \theta(z) \, \rho(\x, z) \, v(\q^2) \, v^*(\q'^2) \, J\left(E,\p-\q\right)J^*\left(E,\p-\q'\right) \notag
     \\&\hspace{5cm}\times  e^{-i(\q-\q')\cdot\x} \, e^{-i\left(Q^-_{p-q}-Q^-_{p-q'}\right)z}\left[1+\u\cdot\boldsymbol{\Gamma}(\q,\q')\right] \, .
 \end{align}

All the dependence on the matter's spatial structure in \eqref{M_1^2 before expanding} comes from the hydrodynamic parameters of the matter model. In our case of the GW potential, these parameters are the source spatial density $\rho(\x,z)$, the Debye mass $\mu^2(\x,z)$, and the three components of the velocity $\u(\x,z)$ and $u_z(\x,z)$. For arbitrary $\x$-dependence in \eqref{M_1^2 before expanding}, the final distribution cannot be simplified further. We proceed by noting that the matter properties are expected to change sufficiently slowly in the hydrodynamic phase, and thus expand in hydrodynamic gradients transverse to the large parton momentum $p_z$, see \cite{Sadofyev:2021ohn}. At the zeroth order, matter is uniform in the transverse plane, and thus the $\x$-integral trivially results in a delta function while the longitudinal $z$-integral is yet to be performed. Then the corresponding contribution to the squared amplitude reads
 \begin{align}
 \label{M1_0^2}
     \langle|M_1|^2\rangle &= \mathcal{C}\,\int_0^L dz \, \int_\q  \, \rho(z) \, |v(\q^2)|^2 \, \left|J\left(E,\p-\q\right)\right|^2\,\left[1+\u\cdot\boldsymbol{\Gamma}(\q)\right] +\mathcal{O}\left(\boldsymbol{\nabla}\right)\,,
 \end{align}
where $\boldsymbol{\Gamma}(\q)\equiv \boldsymbol{\Gamma}(\q,\q)$ and we assume that the matter has a finite longitudinal size $L$.

Given weak dependence on the transverse coordinates, any hydrodynamic variable $G(\x,z)$ can be expanded about $\x=0$, resulting in 
$$
G(\x,z)\simeq G(z)+\x\cdot\boldsymbol{\nabla} G(z)+\cO\left({\boldsymbol \nabla}^2\right)\,.
$$
Then the transverse integral in \eqref{M_1^2 before expanding} can be performed explicitly by noting that
\begin{align}
   \int_{\x} \, \x_\a\, e^{-i \left(\q - \q'\right)\cdot\x} & = i (2\pi)^2 \frac{\partial}{\pa (\q -\q')_\a} \delta^{(2)}\left(\q -\q'\right)\,,
    \label{e:delta-xperp1}
\end{align}
where $\a$ runs over the 2D transverse space. With these replacements, we can express the linear gradient correction to \eqref{M1_0^2} in a compact form
\begin{align}
\label{e:BroadBornM2d}
    &\hspace{-0.5cm}\d\left\langle \left| M_1 \right|^2 \right\rangle =
    -i\mathcal{C}\,\int_0^L dz \int_{\Q,\q_{12}}  \, \d^{(2)}\left(\q_{12}\right) \, \hat{\g}_\a\, \rho(z)  \notag
    \\ 
    & \hspace{0cm} \times \frac{\pa}{\pa \q_{12,\a}}\Bigg\{J\left(E, \p - \Q - \frac{1}{2}\q_{12}\right) J^*\left(E, \p - \Q + \frac{1}{2}\q_{12}\right) v\left(\left(\Q + \frac{1}{2}\q_{12}\right)^2\right) v^*\left(\left(\Q - \frac{1}{2}\q_{12}\right)^2\right)
    \notag \\ & \hspace{0.5cm} \times
    \exp\left[ - i \left(\frac{\u\cdot\q_{12}}{1-u_z}- \, \frac{\q_{12}\cdot(\p-\Q)}{(1-u_z) E }\right)z  \right] 
    \bigg[ 1 
    + \u \cdot \boldsymbol{\Gamma}\left(\Q+ \frac{1}{2}\q_{12},\Q- \frac{1}{2}\q_{12}\right)
    \bigg]\Bigg\} \, , 
\end{align}
where $\Q \equiv \frac{1}{2}\left( \q + \q' \right)$, $\q_{12}\equiv \q - \q'$, and we have integrated the derivative acting on the delta function by parts. All gradient corrections can be derived from the same expression upon variation over the corresponding hydrodynamic variable. We use a shorthand notation, the two-dimensional operator $\hat{\g}_\a\equiv\sum_G\left(\boldsymbol{\nabla}G\cdot\frac{\d}{\d G}\right)_\a$, which should be understood as summed over the hydrodynamic variables. Note that if a $\mu$-dependent cutoff is introduced for the momentum integration, $\hat{\g}$ should, in principle, act on such a cutoff as well.

In order to simplify our considerations, we follow \cite{Sadofyev:2021ohn} and assume that the initial source $J$ and $v$ have at most constant imaginary phases\footnote{Indeed, for arbitrary complex functions $J$ and $v$ one would expect additional contributions proportional to e.g. $J^*(\p)\overset{\text{\tiny$\bm\leftrightarrow$}}{\na}_pJ(\p)$. While non-trivial phases, controlling such structures, are not expected in the commonly used models for the source potential, the initial source is generally less constrained. Here, we will imply this assumption, leaving the case of more general initial distributions aside.}. Therefore, all the terms in the integrand of \eqref{e:BroadBornM2d} --- except for the Landau-Pomeranchuk-Migdal (LPM) phases \cite{Landau:1953um,Migdal:1956tc} --- are even and regular functions of $\q_{12}$ with zero derivative with respect to $\q_{12}$ at zero. This reduces the gradient correction to the amplitude to
\begin{align}
\label{e:BroadBorn10}
    \d\left\langle \left| M_1 \right|^2 \right\rangle &=
    \mathcal{C}\,\int_0^L dz \int_{\q}  \: 
     \hat{\g}_\a  \, \rho(z) \,\left[\left(-\frac{\u^\a}{1-u_z}+\frac{\left(\p-\q\right)^\a}{(1-u_z)E}\right)z\right]\: \notag 
    \\ & \hspace{3cm} \times [v (\q^2)]^2\: |J(E, \p - \q)|^2 \bigg[ 1 + \u \cdot \boldsymbol{\Gamma}(\q) \bigg] \, . 
\end{align}

\subsection{Double Born contribution}
Next we turn to the DB diagram, see Fig.~\ref{f:Broad}, which gives the second contribution to the transverse momentum broadening and corresponds to the double-scattering amplitude:
\begin{align}
    iM_2&=\int \frac{d^4q_1}{(2\pi)^4} \, \frac{d^4q_1}{(2\pi)^4} \, \left[ig t^b_{proj} (2p-q_2)_\nu A^{b\nu}_{\text{ext}}(q_2)\right] \, \left[\frac{i}{(p-q_2)^2+i\epsilon}\right]\notag
    \\ & \hspace{.5cm} \times \left[ig t^a_{proj} (2(p-q_2)-q_1)_\mu A^{a\mu}_{\text{ext}}(q_1)\right] \, \left[\frac{i}{(p-q_1-q_2)^2+i\epsilon}\right] \, J(p-q_1-q_2)\, . 
\end{align}
Unlike the case of the SB diagram, both field insertions in $M_2$ come at the amplitude level, and to the first order in opacity the DB diagram combines with the vacuum amplitude $iM_0=J(p)$. Averaging over the stochastic fields and summing the quantum numbers, we find that
\begin{align}
\label{M2M0W/OInteg}
    \langle M_2M^*_0\rangle &=\mathcal{C} \, \sum_i\,\left(1-2\frac{\u\cdot \p}{E(1-u_{z})}\right)\, \int_{q_1,q_2} \, e^{-i(\q_1+\q_2) \cdot \x_i -i (q_{1z} +q_{2z})z_i} \, v(q_1) \, v(q_2)  \notag
    \\ & \hspace{3cm} \times \frac{2E(1-u_{z})}{(p-q_2)^2+i\epsilon} \,  \frac{2E(1-u_{z})}{(p-q_2-q_1)^2+i\epsilon} \,J(p-q_2-q_1) \, J^*(p) \, ,
\end{align}
where we have used the constraints coming from the external field fixing the temporal components of the two momenta. We further note that, as in the SB case, the dominant contribution to the $q_{1z}$ integral corresponds to the pole at $q_{1z}=-q_{2z}+Q^-_{p-q_1-q_2}-i\epsilon$. Thus, 
\begin{align}
\label{M2M0 d2q1 d3q2}
    \langle M_2M^*_0\rangle &= i \, \mathcal{C} \, \sum_i \, \int_{\q_1,q_2} \, \theta(z_i) \, e^{-i(\q_1+\q_2)\cdot\x_i} \, e^{-iQ^-_{p-q_1-q_2}z_i}\,J(E,\p-\q_1-\q_2) \, J^*(p) \notag
    \\ & \hspace{3cm} \times \left(1-\frac{\u\cdot(2\p-\q_1-\q_2)}{(1-u_{z})E}-\frac{\u\cdot(\q_1+\q_2)}{1-u_{z}}\frac{1}{J}\frac{\partial J}{\partial E}\right) \notag  
    \\ & \hspace{4cm} \times \frac{2E}{1+u_{z}} \, 
    \frac{ v(\tilde{q}_1) \, v(q_2) }{(q_{2z}-Q^+_{p-q_2}-i\epsilon)(q_{2z}-Q^-_{p-q_2}+i\epsilon)} \, ,
\end{align}
where the tilde serves to remind one that $q_{1z}$ and $q_{10}$ are fixed, which introduces a non-trivial $q_{2z}$-dependence into $v(\tilde{q}_{1})$. 

Turning to the $q_{2z}$ integration, one must also note that the Fourier factor is independent of $q_{2z}$, and the residues of the scattering potential poles are no longer suppressed. Therefore, the residues of the all six poles may contribute. The two poles of the second scalar propagator are given by $Q^\pm_{p-q_2}$, while the four poles coming from the scattering potentials read
\begin{subequations}
\begin{align}
    v (\tilde{q}_1): \qquad & \qquad
    \mathcal{P}_1^\pm \equiv 
    Q^-_{p-q_1-q_2} - \frac{u_{z}}{1-u_{z}^2} (\u \cdot \q_{1}) 
    \pm \frac{i}{1-u_{z}^2} R_1  \,  ,
    \\ \label{e:potlpoles1}
    v (q_{2}): \qquad & \qquad
    \mathcal{P}_2^\pm \equiv 
    \frac{u_{z}}{1-u_{z}^2} (\u \cdot \q_{2}) 
    \pm \frac{i}{1-u_{z}^2} R_2 \,  ,
\end{align}
\end{subequations}
where, keeping the expressions compact, we have introduced the shorthand notation
\begin{align}
    \label{e:Rdef}
    R^2 &\equiv (1-u_{z}^2)(\q^2 + \mu^2) - \left(\u \cdot \q \right)^2 \geq 0\,.
\end{align}
Then, introducing the source number density, we can write the full DB contribution as
\begin{align}
    \langle M_2M^*_0\rangle + \text{c.c.} &= i \, \mathcal{C} \, \int_{x,\q_1,\q_2} \,\theta(z) \, \rho(\x,z) \, J(E,\p-\q_1-\q_2) \, J^*(p) \notag
    \\ & \hspace{0.75cm}\times  \left(1-\frac{\u\cdot(2\p-\q_1-\q_2)}{(1-u_{z})E}-\frac{\u\cdot(\q_1+\q_2)}{1-u_{z}}\frac{1}{J}\frac{\partial J}{\partial E}\right) \notag
    \\ & \hspace{1.5cm}\times \, \left[e^{-i(\q_1+\q_2)\cdot\x} \, e^{-iQ^-_{p-q_1-q_2}z} \, \mathcal{I}_{DB} - e^{i(\q_1+\q_2)\cdot\x} \, e^{iQ^-_{p-q_1-q_2}z} \, \mathcal{I}^*_{DB} \right] \,,
    \label{M2withI}
\end{align}
with
\begin{align}
\label{I_DB}
    \mathcal{I}_{DB} &\equiv \int\frac{dq_{2\,z}}{2\pi} \, \frac{2E}{1+u_{z}} \, 
    \frac{ v(\tilde{q}_1) \, v(q_2) }{(q_{2z}-Q^+_{p-q_2}-i\epsilon)(q_{2z}-Q^-_{p-q_2}+i\epsilon)}\,.
\end{align}

At zeroth order in gradients, the hydrodynamic parameters are considered to be constant in the transverse plane, and integrating over $\x$ sets $\q_1=-\q_2\equiv \q$. Therefore, only the imaginary part of the integral $\im\,\mathcal{I}_{DB}(\q,-\q)$ contributes to \eqref{M2withI}. The DB contribution then simplifies to
\begin{align}
\label{M2M0NoGamma}
    \langle M_2M^*_0\rangle + \text{c.c.} &= - \, \mathcal{C} \int_0^L dz \int_{\q} \, \rho(z) \, [v(\q^2)]^2 \,|J(E,\p)|^2 \notag
    \\ & \hspace{1cm} \times  \left[1-\frac{2\, \u\cdot\p}{(1-u_{z})E}  -  \, \frac{\u\cdot\p}{(1-u_z)E} \, \frac{\q^2}{v^2}\frac{\partial v^2}{\partial \q^2}\right]
    +\mathcal{O}\left(\boldsymbol{\nabla}\right) \, .
\end{align}
Introducing a new transverse vector
\begin{align}
\label{Gamma DB}
    \boldsymbol{\Gamma}_{DB}(\q) \equiv -2 \frac{\p}{(1-u_z)E} - \frac{\p}{(1-u_z)E} \frac{\q^2}{[v(\q^2)]^2} \frac{\partial v^2}{\partial \q^2} \, ,
\end{align}
we can re-express \eqref{M2M0NoGamma} in a compact form analogous to \eqref{M1_0^2} 
\begin{align}
    \langle M_2M^*_0\rangle + \text{c.c.} &= - \, \mathcal{C} \int_0^L dz \int_{\q} \, \rho(z) \, [v(\q^2)]^2 \,|J(E,\p)|^2 \, \left[1 + \u \cdot \boldsymbol{\Gamma}_{DB} (\q)\right]
    +\mathcal{O}\left(\boldsymbol{\nabla}\right) \,.
\end{align}

Following the same logic as before, we replace the linear dependence in $\x$ by a derivative of a delta function $\d^{(2)}(\q_1+\q_2)$. Integrating by parts and using shorthand notation, we find the leading gradient corrections to \eqref{M2withI}, which read
\begin{align}
\label{dM2M0general}
    \d\langle M_2M^*_0\rangle + \text{c.c.} &=  \, \mathcal{C} \, \int_0^L dz \int_{\Q,\q_{12}} \: \frac{(2\pi)^2 \d^{(2)}(2\Q)}{2} \, \notag
    \\& \hspace{-2cm}\times \hat{\g}_\a \;\, \frac{\partial}{\partial \Q_\a}\Bigg\{ \,\rho(z)  \,J(E,\p-2\Q) \, J^*(p) 
    \left(1-2\frac{\u\cdot(\p-\Q)}{(1-u_{z})E}-2\frac{\u\cdot\Q}{1-u_{z}}\frac{1}{J}\frac{\partial J}{\partial E}\right) \notag
    \\ & \hspace{3.5cm} \times  \left[ e^{-iQ^-_{p-q_1-q_2}z} \, \mathcal{I}_{DB} + e^{iQ^-_{p-q_1-q_2}z} \, \mathcal{I}^*_{DB} \right] \Bigg\} \,.
\end{align}
where $\Q=\frac{1}{2}(\q_1+\q_2)$ and $\q_{12}=\q_1-\q_2$. The structure of this expression is more involved, and so it is instructive to consider each contribution separately. First, when the derivative acts on the terms in the second line of \eqref{dM2M0general}, the only $\q_{12}$-dependence enters through $\re\, \mathcal{I}_{DB}(\q_{12},0)$. Its angular average scales as $\frac{1}{E}$ and can be explicitly derived.
When the derivative hits the LPM phases, the integrand is proportional to $\im\, \mathcal{I}_{DB}(\q_{12},0)$, and, after averaging over the angles, the corresponding contribution looks like a naive combination of gradient and flow corrections in \cite{Sadofyev:2021ohn}.  
Finally, when the derivative acts on the integral $\mathcal{I}_{DB}$, only its real part contributes. Its eikonal term $\re\, \mathcal{I}^{(0)}_{DB}$ is a regular, even function of $\Q$, and its derivative at $\Q=0$ is zero. In turn, the subeikonal part satisfies $\frac{\partial}{\partial \Q_\a}\,\re\, \mathcal{I}^{(1)}_{DB}(\q_{12},0)=-\frac{\partial}{\partial \Q_\a}\,\re\, \mathcal{I}^{(1)}_{DB}(-\q_{12},0)$. Since the rest of the integrand in \eqref{dM2M0general} is $\q_{12}$-independent, the corresponding contribution to the integral vanishes after angular averaging. Combining these contributions, we find
\begin{align}
    \hspace{-0.5cm}\d\langle M_2M^*_0\rangle + \text{c.c.} &=  \, \mathcal{C} \, \int_0^L dz \int_\q \, \hat{\g}_\a \; \rho(z) \, [v(\q^2)]^2 \, |J(E,\p)|^2   \Bigg\{-\frac{1}{|J|^2} \frac{\partial |J(E,\p)|^2}{\partial \p_\a}  \frac{\pi \, g^4 \, \sqrt{1 - \u^2 - u_z^2}}{4\mu (1-u_z) E}  \frac{\delta^{(2)}(\q)}{[v(\q^2)]^2} \notag
    \\ & \hspace{2cm}+ z\left(\frac{\u_\a}{1-u_z}-\frac{\p_\a}{(1-u_z)E}\right) \,\left[1 +  \, \u \cdot \boldsymbol{\Gamma}_{DB} (\q) \right] 
    \Bigg\}\,.
    \label{dM2M0final}
\end{align}

Finally, one may notice that working with scalar quarks we have to treat the seagull vertex present in the theory and required by consistency. This contribution is commonly omitted in the jet quenching calculations based on scalar QCD, since it is expected to be subeikonal, see e.g. the discussion in \cite{Sadofyev:2021ohn}. However, it may contribute to the leading subeikonal corrections, and cannot be ignored here. It can be easily checked that the corresponding contribution appears only in the presence of transverse anisotropy, and has the same structure as the term in the first line of \eqref{dM2M0final}. This subeikonal term is additionally suppressed by the smallness of the gradients with no length enhancement, and, thus, appears to be subleading to the rest of the terms. In what follows we will neglect such terms, keeping only the length-enhanced subeikonal gradient corrections, and, consequently, neglecting all the diagrams involving the seagull vertex.

\subsection{Final parton distribution and its moments}

The final state parton distribution can be related to the squared amplitude of the process, and is defined by
\begin{align}
   E \frac{d\cN}{d^2p \, dE}\equiv \frac{1}{2(2\pi)^3}  \, \langle |M|^2 \rangle \, ,
\end{align}
where we have chosen $E$ and the two transverse components of the momenta as the independent variables. Introducing the initial parton distribution  $E \frac{d\cN^{(0)}}{d^2p \, dE}\equiv \frac{1}{2(2\pi)^3} \, |J(E,\p)|^2$ and taking the combined effects of flow and gradient corrections together, we find
\begin{align}
\label{dN final}
    E \frac{d\cN}{d^2p \, dE} &=  E \frac{d\cN^{(0)}}{d^2p \, dE} \notag
    \\  
    &\hspace{-1cm} +\mathcal{C} \, \int_0^L dz \int_\q \, \Bigg\{ \left[1 - \hat{\g}_\a \, \frac{(\u E-\p+\q)_\a \, z}{(1-u_z)E} \right] \, [1 + \u \cdot \boldsymbol{\Gamma}(\q)] \, E \frac{d\cN^{(0)}}{d^2(p-q) \, dE} \, \notag
    \\ 
    & \hspace{0cm} -\left[1 - \hat{\g}_\a \, \frac{(\u E-\p)_\a \, z}{(1-u_z)E} \right] \, [1 + \u \cdot \boldsymbol{\Gamma}_{DB}(\q)] \, E \frac{d\cN^{(0)}}{d^2p \, dE}\Bigg\} \, \rho(z) \, [v(\q^2)]^2 \, ,
\end{align}
where the terms appearing at the second subeikonal order should be neglected. Integrating the final-state distribution over the transverse momentum, one can show that the number of energetic partons is unchanged as long as the initial distribution falls fast enough at infinity, as required by unitarity.

It is instructive to consider how a particular ensemble of partons is modified by a flowing anisotropic distribution of matter. To do so, we focus on a narrow initial distribution parametrized as
\begin{align}
\label{dN0 for moments}
    E \frac{d\cN^{(0)}}{d^2p \, dE} = \frac{f(E)}{2 \pi w^2} e^{-\frac{\p^2}{2 w^2}} \, .
\end{align}
Starting with the corresponding family of partons in the initial state, one may understand how their distribution changes by focusing on the leading moments of the final state distribution, defined as
\begin{align}
    \langle \p_{\a_1}...\p_{\a_n} \rangle \equiv \frac{\int_\p \, (\p_{\a_1}...\p_{\a_n}) \, E \frac{d\cN}{d^2p \, dE}}{\int_\p \, E \frac{d\cN^{(0)}}{d^2p \, dE}} = (2\pi)^2 \int_\p \, \frac{(\p_{\a_1}...\p_{\a_n})}{f(E)} \, E \frac{d\cN}{d^2p \, dE} \, .
\end{align}

The process of jet-transverse-momentum broadening is often quantified with the so-called jet quenching parameter, which is argued to control many of the related processes. Thus, we start by computing the quadratic moment given by
\begin{align}
    \langle \p^2 \rangle  = 2w^2 + \mathcal{C} \, \int_0^L dz \, \left[1 - z\,\hat{\g} \cdot \frac{\u}{1-u_z} \right] \rho(z) \, \int_\q \,  \q^2 \, [v(\q^2)]^2  \, ,
\end{align}
where the first term corresponds to the vacuum part of \eqref{dN final} and is fixed by the width of the initial distribution. Varying this moment with the medium length $L$ --- understood as the path length travelled by the parton --- we find  
\begin{align} \label{e:qhat_modified}
\hat{q}(L)=\left[1 - L\,\hat{\g} \cdot \frac{\u}{1-u_z} \right]\hat{q}_0(L)\,,
\end{align}
where $\hat{q}_0(L)=\mathcal{C} \rho(L) \, \int_\q \, \q^2 \, [v(\q^2)]^2 $ corresponds to the limit of static transversely homogeneous matter, and all hydrodynamic variables and their gradients should be understood as functions of $L$. Thus, we can see that the interplay between the transverse medium structure and flow modifies the even moments of the final state distribution, including the fundamental jet quenching parameter, at the leading eikonal order. Moreover, this modification is length-enhanced, and consequently, expected to be the dominant effect of the medium's evolution. This is one of the main results of this work.

Following \cite{Sadofyev:2021ohn,Andres:2022ndd,Barata:2022krd}, we also consider the leading odd moments of the final distribution, which vanish in the case of static and isotropic matter (for isotropic initial distribution) but generated by flow and gradient effects.  One readily finds that
\begin{align}
    \langle \p_\a \rangle &=  - \frac{1}{2}\mathcal{C} \int_0^L dz \,  \left[1 - z\,\hat{\g} \cdot \frac{\u}{1-u_z} \right] \, \rho(z)\notag\\
    &\hspace{2cm}\times \frac{\u_\a}{(1-u_z)E} \, \int_\q \q^2 \, \left[ E \frac{f'(E)}{f(E)} +\q^2 \, \frac{\partial}{\partial \q^2} \right] \, [v(\q^2)]^2 
    \, ,
\end{align}
where we recognize the same re-scaling factor as in the case of even moments combined with the averaged transverse momentum obtained in \cite{Sadofyev:2021ohn,Andres:2022ndd}. 

As shown in \cite{Sadofyev:2021ohn, Barata:2022krd}, the directional gradient effects appear only in higher odd moments of the final parton distribution. Focusing on the cubic moment, we find
\begin{align}
    \langle \p_\a \, \p^2 \rangle &= \mathcal{C} \int_0^L dz \int_\q \, \Bigg\{ 2w^2\,z\hat{\g}_\a \,  \frac{\q^2 }{(1-u_z)E} - \frac{1}{2}\left[1 - z\,\hat{\g} \cdot \frac{\u}{1-u_z} \right] \, \frac{\u_\a}{(1-u_z)E} \,  \notag
    \\ & \hspace{1cm} \times \q^2\left[8 w^2  +  (10 w^2 + \q^2) \,\q^2\frac{\pa}{\pa \q^2} + \left(4 w^2 +\q^2\right) \, E\frac{f'(E)}{f(E)}\right] \Bigg\} \, \rho(z) \, [v(\q^2)]^2 \, ,
\end{align}
which agrees with the results of \cite{Sadofyev:2021ohn,Barata:2022krd} in the limit of static matter at first order in opacity, and can be compared to the results of \cite{Sadofyev:2021ohn} in the homogeneous limit at $w=0$.

\section{Medium-induced gluon spectrum}\label{SecIII:Branching}

In this section, we will derive the medium-induced radiation spectrum in an evolving inhomogeneous nuclear matter at the first order in opacity, and discuss the resulting modifications of the energy loss. This consideration is technically more involved, and here we will focus on two particular types of the medium evolution effects. First, we will study how the transverse gradients of $\m$ and $\rho$ modify the full gluon spectrum in the absence of any flow, at the first order in opacity, but in full kinematics, and thereby extending the results of \cite{Sadofyev:2021ohn,Barata:2023qds}. Second, we will search for the mixed flow-gradient effects analogous to the multiplicative modification in \eqref{e:qhat_modified}, keeping only the leading terms in the eikonal expansion. For notational compactness, we will put the two types of gradient corrections together in intermediate equations, although one should note that the subeikonal terms sourced by the transverse flow (and their combinations with the leading gradient corrections) are omitted\footnote{It should be mentioned here that in this limit the class of diagrams involving the seagull or four-gluon vertices can be freely neglected. Indeed, they contribute as gradient corrections, and appear only at the first subeikonal order, lacking any length enhancment in the absence of the transverse flow.}. In the absence of $u_z$ gradients, the longitudinal flow effects can be obtained with the corresponding boost, and for simplicity we will set $u_z=0$.

Before turning to details, we should note that the external field \eqref{Amu} has been obtained from the classical field equation in the Lorenz gauge. However, it is convenient to impose a stronger gauge condition, additionally requiring $n\cdot A=0$. In this gauge, only the two physical gluon polarizations appear, and the consideration simplifies. This constraint is compatible with the form of \eqref{Amu}, and one may choose $n^\m$ to be any vector transverse to the velocity $u^\m$. Here, we will use $n_\mu=(0,\,\boldsymbol{0},\,1)$ since $A^z_{\text{ext}}=0$ for $u_z=0$. In this gauge, the propagator of a gluon is given by 
\begin{align}
    G_{\mu\nu}(k)=\frac{-iN_{\mu\nu}(k)}{k^2+i\epsilon} \, ,
\end{align}
where the numerator is transverse to the gauge vector $n$ and gluon momentum $k$
\begin{align}
     N_{\mu\nu}(k)=g_{\mu\nu}+n^2 \frac{k_\mu k_\nu}{(k\cdot n)^2-k^2 n^2} + k^2 \frac{n_\mu n_\nu}{(k\cdot n)^2-k^2 n^2}
    -(k\cdot n)\frac{k_\mu n_\nu + k_\nu n_\mu}{(k\cdot n)^2 -k^2 n^2} \, .
\end{align}
The polarization vector now satisfies $k\cdot\epsilon(k)=n\cdot\epsilon(k)=0$, and can be parametrized as
\begin{align}
    \epsilon^{*}_\mu(k) = \left(\frac{\beps\cdot\k}{\o}, \,\beps, \,0 \right)
\end{align}
where $\beps$ is the transverse polarization\footnote{The transverse polarization vectors are functions of $k$, since the sum over polarizations should be properly normalized. However, up to the higher subeikonal corrections, we find that $\sum_{\l} \beps^\l_\a \beps^\l_\b = \d_{\a\b} +\cO\left(\frac{1}{\o^2}\right)$, with $\a$ and $\b$ running in the 2d transverse space.}, $k=(\o,\k,k_z)$, and we keep the polarization index implicit.

\begin{figure}[t]
    \centering
    \vspace*{-8mm}
	\includegraphics[width=0.7\textwidth]{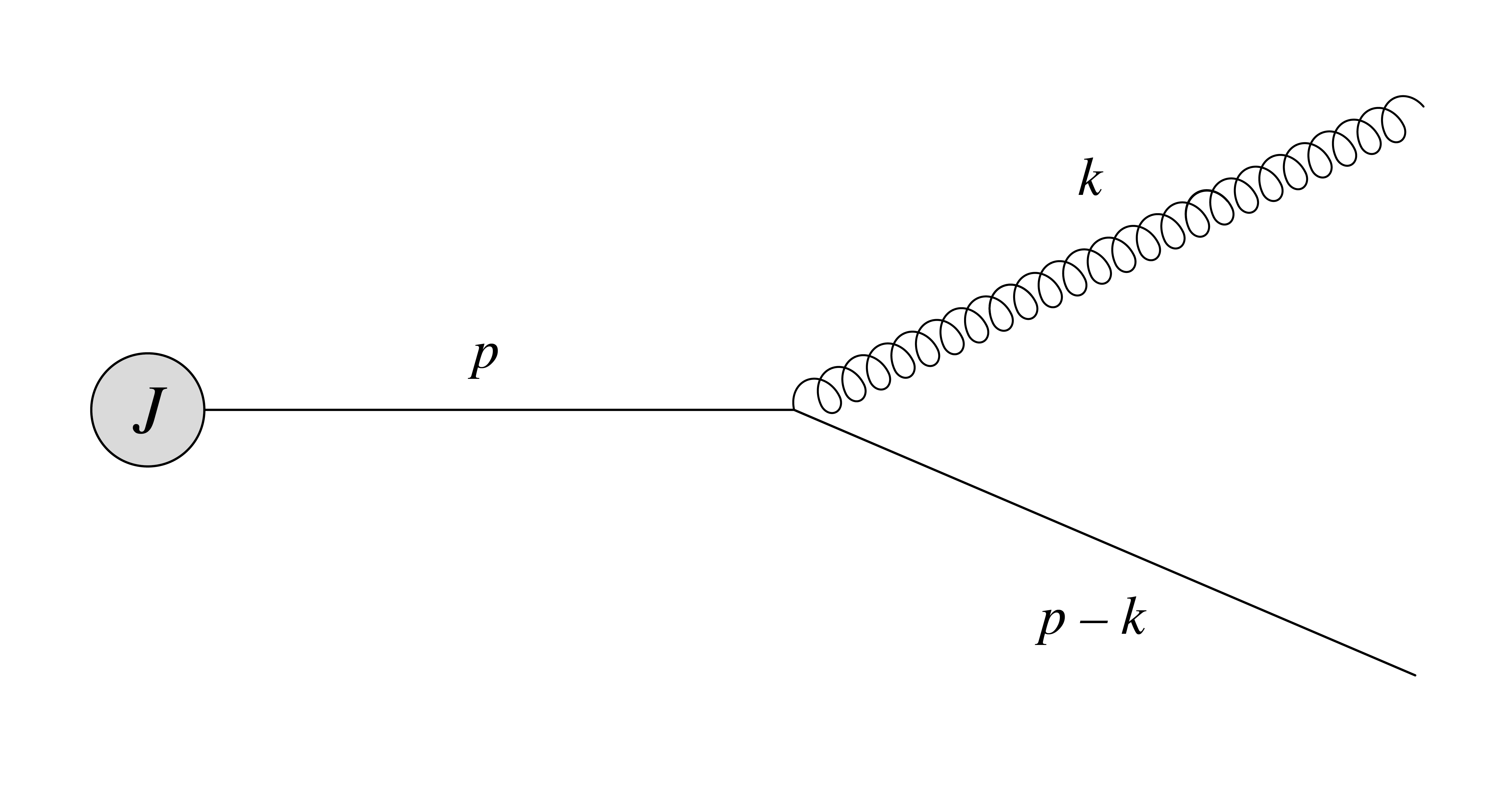}
	\vspace*{-10mm}\caption{The elementary vacuum splitting, corresponding to the zeroth order in opacity expansion.}
	\label{f:vacuum}
\end{figure}

\begin{figure}[t!]
    \centering
    \includegraphics[width=1\textwidth]{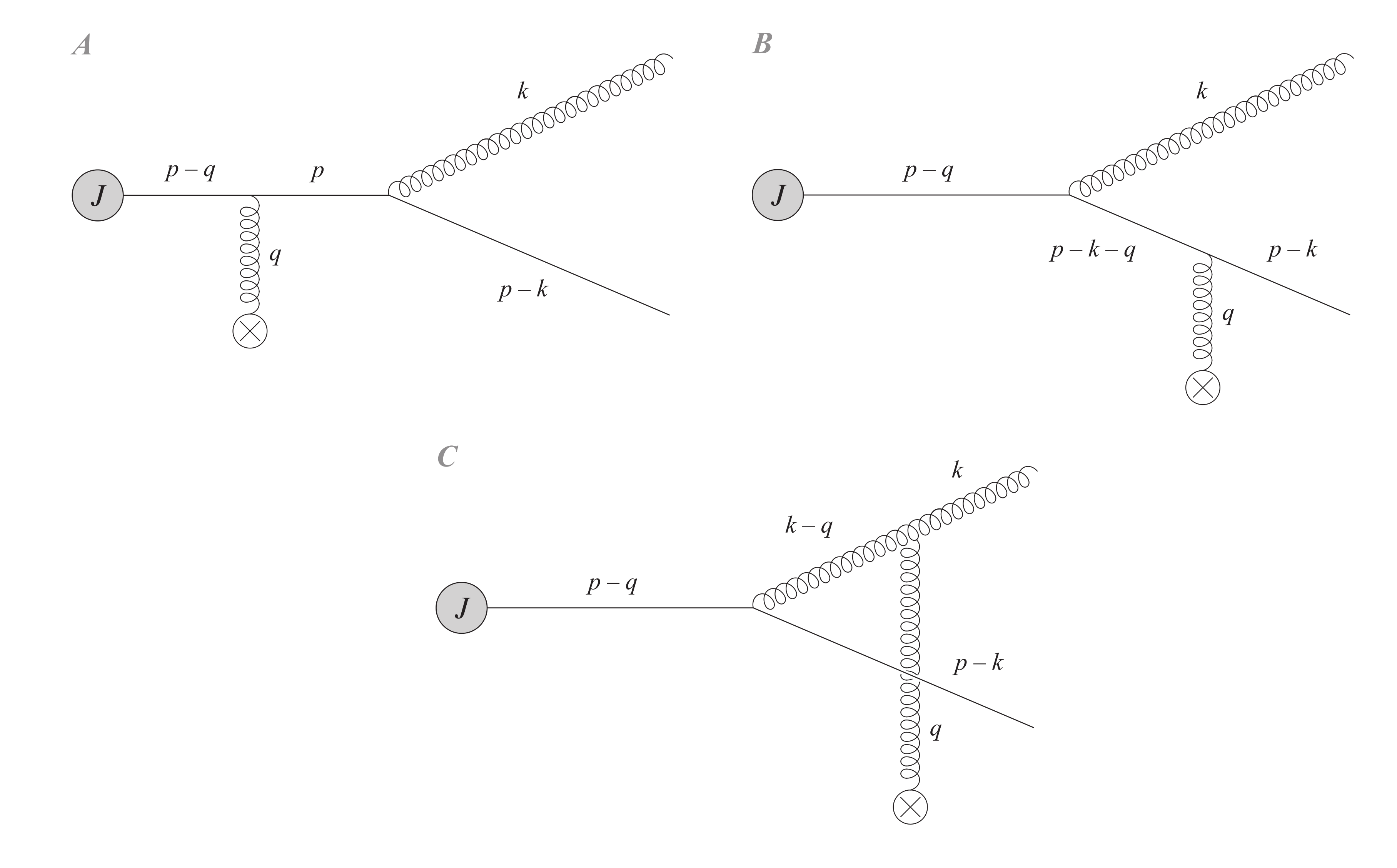}
    \vspace*{-10mm}
	\caption{The three single-Born diagrams contributing to the medium-induced radiation at the first order in opacity.}
	\label{f:matter_SB}
\end{figure}

Following the notation of \cite{Sadofyev:2021ohn,Barata:2023qds}, we consider the leading perturbative contribution to the medium-induced branching of the initial energetic quark to an on-shell quark of momentum $(p-k)_\mu$ and an on-shell gluon of momentum $k_\mu$, working at the leading eikonal order but to all orders in LPM phases. 
We also rely on the broad source approximation, assuming that the dependence of the initial source on the transverse momenta is sufficiently weak. 
Under the aforementioned assumptions, the vacuum emission amplitude (see Fig.~\ref{f:vacuum}) corresponding to the zeroth order in opacity, reads
\begin{align}
    i\cR_0=\left[i\, g\, t^r_{proj} \, (2p-k)_\mu \,\epsilon^{*\mu}(k)\right] \, \frac{i}{p^2+i\epsilon}  \, J(p) \simeq - g\, t^r_{proj} \, \frac{2(1-\text{x}) \, \beps \cdot (\k-\text{x}\p)}{(\k - \text{x}\p)^2} \, J(p) \, ,
\end{align}
where $t^r_{proj}$ comes from the emission vertex, $r$ is the color of the final gluon, $\text{x}=\frac{\o}{E}$ is its energy fraction, and one may recognize the (light-front) wave function of the splitting $\psi(\text{x}, \k-\text{x}\p)= \frac{2(1-\text{x}) \, \beps \cdot (\text{x}\p - \k)}{(\k - \text{x}\p)2}$ which controls the branching probability, see e.g. the discussion in \cite{Sievert:2018imd}. In turn, at the first order in opacity $N=1$, there are 9 diagrams contributing to the amplitude of the medium-induced branching shown in Fig.~\ref{f:matter_SB} and \ref{f:matter_DB}, and we will consider them one-by-one in what follows. 

%
%

\subsection{Single Born contributions}

\subsubsection{The amplitude level} 

Let us start by computing the simplest contribution to the medium-induced branching -- $\cR_A$, which corresponds to the initial state scattering. This amplitude reads
\begin{align}
\label{RA beginning}
    i\cR_A & =\int\frac{d^4q}{(2\pi)^4}\, [i\,g\, t^r_{proj} \,(2p-k)_\nu \,\epsilon^{*\nu}(k)] \, \frac{i}{p^2+i\epsilon} \, 
    [i \, t^a_{proj} \, (2p-q)_\mu \,g A^{a\mu}(q)] \, \frac{i}{(p-q)^2+i\epsilon} \, J(p-q) \notag
    \\ & \hspace{0.5cm} =\frac{2(1-\text{x})\,\beps \cdot (\k - \text{x}\p)}{(\k - \text{x}\p)^2} \, \sum_i \, t^r_{proj}  t^a_{proj}t^a_i \, \int_q \, e^{-i(\q_i\cdot\x_i + q_z z_i)} \frac{(2\,g\,E)\,v(q)}{(p-q)^2+i\epsilon} \, J(p-q) \, , 
\end{align}
where we have explicitly used the constraint on $q_0$. Turning to the $q_z$-integration, we again assume that the residues of the scattering potential poles are exponentially suppressed. In turn, the poles coming from the quark propagator are given by
\begin{subequations}
\begin{align}
    \cQ_{p-q}^+ &\simeq 2E \left(1- \frac{\q\cdot\u}{2E}\right) \,
    , 
    \\  \cQ_{p-q}^- &\simeq \q\cdot\u - \frac{(1-\text{x})\k^2 + \text{x}(\p-\k)^2-\text{x}(1-\text{x})(\p-\q)^2}{2\text{x}(1-\text{x})E} \, ,
\end{align}
\end{subequations}
while the initial source is assumed to result in no new poles. Thus, collecting all the relevant contributions, we can write the initial-state scattering amplitude as 
\begin{align}
    i\cR_A&= - ig \, \frac{2(1-\text{x})\,\beps \cdot (\k - \text{x}\p)}{(\k - \text{x}\p)^2} \, \sum_i \, t^r_{proj}  t^a_{proj}t^a_i\notag
    \\
    &\hspace{2cm}\times \int_\q \, \theta(z_i) \, e^{-i\q\cdot\x_i}\,e^{-i\cQ^-_{p-q} \, z_i} \, v(\q^2) \, J(E,\p-\q) \, .
\end{align}
%

%
%

The next diagram corresponds to the final-state scattering process in the quark line, and its amplitude reads 
\begin{align}
    i\cR_B &= \int \frac{d^4q}{(2\pi)^4}\, [i\, t^a_{proj}\, (2(p-k)-q)_\mu \, gA_{ext}^{a\mu}] \, \frac{i}{(p-k-q)^2+i\epsilon} \,\notag
    \\ & \hspace{2cm} \times [i \,g \,t^r_{proj}  (2(p-q)-k)_\nu \, \epsilon^{*\nu}(k)] \, \frac{i}{(p-q)^2+i\epsilon} \, J(p-q) \, ,
\end{align}
which, after both $q_0$ and $q_z$ integrals are evaluated, can be rewritten as 
\begin{align}
    i\cR_B &= i g \, \sum_i t^a_{proj}t^a_i t^r_{proj} \int_\q \, \theta(z_i) \, e^{-i\q\cdot\x_i} \, \frac{2(1-\text{x}) \beps \cdot (\k - \text{x}(\p-\q))}{(\k - \text{x}(\p-\q))^2}  \,  \notag
    \\ & \hspace{2cm} \times
    \left[e^{-i \cQ^-_{p-q} \, z_i} - e^{-i \cQ^-_{p-k-q} \, z_i} \right] \, v(\q^2) \, J(E,\p-\q) \, ,
\end{align}
where we have used the explicit form of the poles of the second quark propagator
\begin{subequations}
\begin{align}
     \cQ_{p-k-q}^+ &\simeq 2(1-\text{x})E \left(1- \frac{\q\cdot\u}{2(1-\text{x})E}\right) \,  , 
     \\  \cQ^-_{p-k-q} &\simeq \q \cdot \u  +\frac{(\p-\k-\q)^2 - (\p-\k)^2}{2(1-\text{x})E}\, .
\end{align}
\end{subequations}

Turning to the last SB contribution in the amplitude, corresponding to the final-state scattering in the emitted gluon line, we write
\begin{align}
    i\cR_C &= \int \frac{d^4 q}{(2\pi)^4} \, i(2p-k-q)_\mu \, t^a_{proj} \, \frac{-i N^{\mu\nu}(k-q)}{(k-q)^2+i\epsilon} \,\Gamma^{abc}_{\nu\alpha\rho}(k-q,q,-k) \notag\,
    \\ &\hspace{3.5cm} \times  
    gA^{b \alpha}(q) \, \epsilon^{*\rho}(k) \, \frac{i}{(p-q)^2+i\epsilon}  \, J(p-q) \, ,
\end{align}
where $\Gamma^{abc}_{\nu\alpha\rho}(k-q,q,-k)$ is the three-gluon vertex defined as 
\begin{align}
    \Gamma^{abc}_{\mu\nu\rho}(k,p,q) = g\,f^{abc}\,[g_{\mu\nu} (k-p)_{\rho} + g_{\nu\rho} (p-q)_\mu + g_{\rho\mu} (q-k)_\nu] 
\end{align}
with all the momenta going towards the vertex. Since $N^{\mu\nu}$, $\epsilon^*_\mu$, and $A^{a\mu}_{\text{ext}}$ are all transverse to $n_\mu=(0, \,\boldsymbol{0}, \,1)$ by construction, the product of these three objects and the three-gluon vertex is independent of the $z$-components of momenta, resulting in no new poles. Performing both the $q_0$ and $q_z$ integrals, we can write $\cR_C$ as
\begin{align}
     i\cR_C &= g \, \sum_i \, f^{abc} t^a_{proj} t^b_i \int_\q \,  \theta(z_i)\,e^{-i\q \cdot \x_i} \, \frac{2(1-\text{x}) \, \beps \cdot (\k - (1-\text{x})\q -\text{x}\p)}{(\k - (1-\text{x})\q - \text{x}\p)^2} \, v(\q^2) \, \notag
     \\ & \hspace{2.5cm} \times\left[e^{-i \cQ^-_{p-q} \, z_i}- e^{-i \cQ^-_{k-q} \, z_i}\right] \,  J(E,\p-\q) \, ,
\end{align}
where the poles of the gluon propagator have been used
\begin{subequations}
\begin{align}
     \cQ_{k-q}^+ &\simeq 2\text{x}E \left(1- \frac{\q\cdot\u}{2\text{x}E}\right) \,  , 
     \\  \cQ^-_{k-q} &\simeq \q \cdot \u + \frac{(\k-\q)^2-\k^2}{2\text{x}E}\, .
\end{align}
\end{subequations}

\subsubsection{SB contributions to the squared amplitude}

Now we are in position to obtain the full SB contribution to the amplitude squared. As done in the case of jet broadening, we have to average over initial and sum over final quantum numbers, as well as perform the medium averages.  

Let us start by squaring $\cR_A$, which after averaging but before integrating over the transverse coordinates results in
\begin{align}
    \langle| \cR_A |^2\rangle &=  \frac{C_F^2}{2N_c}\, g^2 \, \frac{4(1-\text{x})^2}{(\k - \text{x}\p)^2} \, \int_{x,\q,\Bar{\q}} \, \theta(z) \, \rho(\x,z) \, 
    e^{-i(\q-\bar{\q}) \cdot \x}   \notag
    \\ & \hspace{1cm} \times e^{-i(\cQ^-_{p-q}-\cQ^-_{p-\bar{q}}) \, z}
    \, v(\q^2) \, v(\bar{\q}^2)\,  J(E,\p - \q) J^*(E,\p - \bar{\q}) \, ,
\end{align}
where the color factors combine into an overall multiple $\frac{C_F^2}{2N_c}$. In the case of a transversely homogeneous matter, corresponding to the zeroth order in the gradient expansion, we can readily evaluate the $\x$-integral. It results in a constraint $(2\pi)^2 \d^{(2)}(\q-\bar{\q})$, and integrating over one of the momenta the standard form of the leading contribution is obtained. In turn, treating the gradient corrections as in the broadening case, one may note that only the LPM phases contribute to the momentum derivative at $\q-\bar{\q}=0$. Thus, up to the first order in gradients and in the eikonal limit for the flow induced terms, the corresponding contribution to the amplitude squared reads
\begin{align}
    \langle |\cR_A^2|\rangle &= \frac{C^2_F }{N_c} \, g^2 \, \int_0^L dz \int_\q \, \frac{2(1-\text{x})^2}{(\k-\text{x}\p)^2}\left[1- \hat{\g} \cdot \left(\u - \frac{\p-\q}{E}\right) \, z\right] \,    \notag
    \\ & \hspace{8cm} \times \rho(z) \, [v(\q^2)]^2 \, \left|J(E,\p-\q)\right|^2 \, .
\end{align}
It should be stressed that if the subeikonal flow corrections, scaling as $\mathcal{O}\left(\frac{\perp}{E}\right)$, were added here, they would mix with the eikonal gradient terms, scaling as $\mathcal{O}\left(|\u| z\right)$, giving contributions of order $\mathcal{O}\left(|\u|^2\frac{\perp}{E}z\right)$, which are formally of the same smallness as the subeikonal (but length enhanced) gradient corrections, scaling as $\mathcal{O}\left(\frac{\perp}{E}z\right)$. These are the terms which we omit in our consideration, focusing on the two limits described above -- (i) keeping the terms scaling as $\mathcal{O}\left(\frac{\perp}{E}z\right)$ but with no transverse flow, (ii) keeping only the eikonal gradient corrections in the presence of a transverse flow.

Squaring $\cR_B$, and averaging the result over the background field configurations, we find
\begin{align}
    \hspace{-0.5cm}\langle |\cR_B|^2 \rangle &= \frac{C_F^2}{2N_c} \, g^2 \,  \int_{x,\q,\Bar{\q}} \, \theta(z) \, \rho(\x,z) \, e^{-i (\q-\bar{\q}) \cdot \x} \, \notag
    \\ & \hspace{-1cm} \times \frac{4(1-\text{x})^2 \, (\k - \text{x}(\p - \q)) \cdot (\k - \text{x}(\p - \bar{\q}))}{(\k - \text{x}(\p - \q))^2 (\k - \text{x}(\p - \bar{\q}))^2} \, v(\q) \, v(\bar{\q}) \, J(E,\p-\q)\, J^*(E,\p-\bar{\q}) \,  \notag
    \\ & \hspace{-0.5cm} \times  \left[e^{-i (\cQ^-_{p-q}-\cQ^-_{p-\bar{q}}) \, z} + e^{-i (\cQ^-_{p-k-q}- \cQ^-_{p-k-\bar{q}}) \, z} - e^{-i  (\cQ^-_{p-q} - \cQ^-_{p-k-\bar{q}}) \, z} - e^{i (\cQ^-_{p-\bar{q}}-\cQ^-_{p-k-q}) \, z} \right] \, .
\end{align}
Accounting for the fact that this SB contribution involves two LPM phases, we find that up to the first order in gradients the corresponding contribution to the amplitude squared reads 
\begin{align}
     \langle |\cR_B|^2\rangle &= \frac{C^2_F}{N_c} \,g^2\, \int_0^L dz \int_\q \, \frac{4 (1-\text{x})^2}{(\mathbf{k}-\text{x}(\p-\q))^2}   \left[1- \hat{\g} \cdot \left(\u -\frac{\p-\q}{2E} - \frac{\p-\k-\q}{2(1-\text{x})E}\right) z\right] \, \notag
    \\ & \hspace{1cm} \times \left[1 -\cos\left(\frac{(\k-\text{x}(\p-\q))^2 }{2\text{x}(1-\text{x})E} \, z\right) \right] \, \rho(z) \, [v(\q^2)]^2 \, \left|J(E,\p-\q)\right|^2 \, .
\end{align}

The last SB contribution $\cR_C$, corresponding to the final-state scattering process in the emitted gluon line, also involves two LPM phases, and, after squaring and averaging, it reads
\begin{align}
    \hspace{-0.5cm}\langle |\cR_C|^2 \rangle &= \frac{C_F C_A}{2 N_c} \, g^2 \,\int_{x,\q,\Bar{\q}} \theta(z)\, \rho(\x,z) \, e^{-i(\q-\bar{\q}) \cdot \x} \,  \notag
    \\ & \hspace{-1cm} \times \frac{4(1-\text{x})^2\, (\k - (1-\text{x})\q - \text{x}\p) \cdot (\k - (1-\text{x})\bar{\q} - \text{x}\p)}{(\k - (1-\text{x})\q - \text{x}\p)^2 (\k - (1-\text{x})\bar{\q} - \text{x}\p)^2} \,  v(\q) \,   v(\bar{\q}) \, J(E,\p-\q)  J^*(E,\p-\bar{\q}) \notag
    \\ &  \hspace{-0.5cm} \times \left[e^{-i (\cQ^-_{p-q}-\cQ^-_{p-\bar{q}}) \, z} + e^{-i (\cQ^-_{k-q}-\cQ^-_{k-\bar{q}}) \, z} - e^{-i (\cQ^-_{p-q}-\cQ^-_{k-\bar{q}}) \, z} - e^{i (\cQ^-_{p-\bar{q}}-\cQ^-_{k-q}) \, z}\right] \, .
\end{align}
As in the cases of $\cR_A$ and $\cR_B$, the only non-vanishing contribution to the linear gradient correction comes from the derivative acting on the phases, and we find 
\begin{align}
    \langle |\cR_C|^2 \rangle &= C_F \, g^2 \int_0^L dz \int_{\q} \, \frac{4(1-\text{x})^2}{(\k - (1-\text{x})\q-\text{x}\p)^2} \, \left[1 - \hat{\g} \cdot \left(\u -\frac{\p-\q}{2E}-\frac{\k-\q}{2\text{x}E}\right)\, z\right] \notag
    \\ & \hspace{1cm} \times  \left[1 -\cos\left(\frac{(\k-(1-\text{x})\q-\text{x}\p)^2 }{2\text{x}(1-\text{x})E} \, z\right) \right] \, \rho(z) \, [v(\q^2)]^2 \, \left|J(E,\p-\q)\right|^2 \, .
\end{align}

The three SB interference terms can be computed following the very same procedure -- the $\x$-dependence of the matter parameters is expressed through the corresponding momentum derivatives. However, the light-front wave functions are structurally different in the direct and conjugated amplitudes, and that results in a new class of gradient corrections. For instance, the interference term $\langle \cR_A \cR^*_B \rangle$ can be written as
\begin{align}
    \langle \cR_A \cR^*_B \rangle + \text{c.c.} &= \frac{C_F}{ N_c^2} \, g^2 \, \int_0^L dz \int_\q \, \frac{2(1-\text{x})^2 \, (\k-\text{x}\p) \cdot (\k - \text{x}(\p-\q))} {(\mathbf{k} - \text{x}\p)^2 (\mathbf{k} - \text{x}(\p-\q))^2} \notag
    \\ & \hspace{-2.5cm} \times \Bigg\{  \left[1-  \hat{\g} \cdot \u \, z\right]  \, \left[1 -\cos\left( \frac{(\k - \text{x}(\p-\q))^2}{2\text{x}(1-\text{x})E}\, z \right) \right] \, \notag
    \\ & \hspace{-1.5cm} +  \hat{\g} \cdot \left( \frac{\p - \q}{E} z - \left( \frac{\p - \q}{2E}\, z + \frac{\p-\k-\q}{2(1-\text{x})E}\, z \right)  \cos\left( \frac{(\k - \text{x}(\p - \q) )^2}{2\text{x}(1-\text{x})E}\, z \right) \right)\, \notag
    \\ & \hspace{-1.5cm} - \text{x} \,  \hat{\g} \cdot \left( \frac{\k - \text{x}(\p - \q )}{(\k-\text{x}(\p-\q))^2} - \frac{\k-\text{x}\p}{2 \, (\k-\text{x}\p) \cdot (\k - \text{x}(\p-\q))}\right) \, \notag
    \\ & \hspace{1cm} \times \sin\left( \frac{(\k-\text{x}(\p-\q))^2}{2\text{x}(1-\text{x})E}\, z \right)\Bigg\} \, \rho \, [v(\q^2)]^2 \,  \left|J(E,\p-\q)\right|^2 \, ,
\end{align}
where the sine phase structure has no explicit length enhancement since it involves no momentum derivatives of the LPM phases. In turn, the two other combinations read
\begin{align}
    \langle \cR_A \cR^*_C \rangle + \text{c.c.} & = - C_F \, g^2 \int_0^L dz \int_\q \, \frac{2(1-\text{x})^2 \, (\k - \text{x}\p) \cdot (\k - (1-\text{x})\q-\text{x}\p)}{(\k-\text{x}\p)^2 (\k -(1-\text{x})\q-\text{x}\p)^2} \, \notag
    \\ & \hspace{-2.5cm} \times \Bigg\{  \left[1- \hat{\g} \cdot \u \, z\right]   \left[1 -\cos\left(\frac{(\k - (1-\text{x})\q-\text{x}\p)^2}{2 \text{x} (1-\text{x}) E} \, z\right) \right] \, \notag
    \\ & \hspace{-1.5cm} +  \hat{\g} \cdot \left[\frac{\p-\q}{E}z - \left( \frac{\p-\q}{2E} z + \frac{\k-\q}{2\text{x}E} z\right) \cos\left(\frac{\k - (1-\text{x})\q-\text{x}\p^2}{2 \text{x} (1-\text{x}) E} \, z\right) \right] \, \notag
    \\ & \hspace{-1.5cm} + (1-\text{x}) \, \hat{\g} \cdot \left( \frac{\k -(1-\text{x})\q -\text{x}\p}{(\k -(1-\text{x})\q -\text{x}\p)^2} - \frac{\k - \text{x}\p }{2 \, (\k - \text{x}\p) \cdot (\k - (1-\text{x})\q-\text{x}\p)}\right) \, \notag
    \\ & \hspace{1cm}  \times \sin\left(\frac{(\k -(1-\text{x})\q - \text{x}\p)^2}{2\text{x}(1-\text{x})E} \, z\right) \Bigg\} \, \rho \, [v(\q^2)]^2 \, \left|J(E,\p-\q)\right|^2 \, ,
\end{align}
and 
\begin{align}
    \langle \cR_B \cR^*_C \rangle + \text{c.c.} &=  - C_F \, g^2 \int_0^L dz \int_\q \, \frac{2(1-\text{x})^2 \, (\k - \text{x}(\p-\q)) \cdot (\k - (1-\text{x})\q -\text{x}\p)}{(\k-\text{x}(\p-\q))^2 (\k-(1-\text{x})\q - \text{x}\p)^2} \, \notag
    \\ & \hspace{-2.5cm} \times \Bigg\{ \left[1 - \hat{\g} \cdot \u \, z\right]  \bigg[1 + \cos\left( \frac{\q \cdot (2 \k - (1-2\text{x})\q- 2\text{x} \p)}{2\text{x}(1-\text{x})E} \, z\right)   \notag
    \\ & \hspace{1cm} - \cos\left(\frac{(\k - (1-\text{x})\q -\text{x}\p)^2}{2\text{x}(1-\text{x})E}\, z\right) - \cos\left(\frac{(\k-\text{x}(\p-\q))^2}{2\text{x}(1-\text{x})E}\, z\right)\bigg] \, \notag
    \\ & \hspace{-1.5cm} + \hat{\g} \cdot \, \notag \bigg[ \frac{\p-\q}{E} z - \left(\frac{\p-\q}{2E} z + \frac{\p-\k-\q}{2(1-\text{x})E} z \right) \cos\left(\frac{(\k-\text{x}(\p-\q))^2}{2\text{x}(1-\text{x})E}\, z\right) \, \notag
    \\ & \hspace{0cm} - \left(\frac{\p-\q}{2E} z + \frac{\k-\q}{2\text{x}E} z \right) \cos\left(\frac{(\k - (1-\text{x})\q -\text{x}\p)^2}{2\text{x}(1-\text{x})E}\, z\right) \, \notag
    \\ & \hspace{1.5cm} + \left(\frac{\p-\k-\q}{2(1-\text{x})E} z + \frac{\k-\q}{2\text{x}E} z \right) \cos\left( \frac{\q \cdot (2 \k - (1-2\text{x})\q- 2\text{x} \p)}{2\text{x}(1-\text{x})E} \, z\right)  \bigg]  \, \notag
    \\ & \hspace{-1.5cm} -\hat{\g} \cdot\Bigg( (1-\text{x}) \, \frac{\k - (1-\text{x})\q -\text{x}\p}{(\k - (1-\text{x})\q-\text{x}\p)^2} - \frac{\k - \text{x}\p}{2 \, (\k - \text{x}(\p-\q)) \cdot (\k - (1-\text{x})\q -\text{x}\p)} \, \notag
    \\ & \hspace{0cm}  + \text{x} \, \frac{\k - \text{x}(\p - \q)}{(\k - \text{x} (\p-\q))^2}\Bigg) \, \bigg[\sin\left(\frac{(\k - \text{x}(\p-\q))^2}{2\text{x}(1-\text{x})E}\, z\right) \notag
    \\ & \hspace{1.5cm}  -\sin\left(\frac{(\k - (1-\text{x}) \q - \text{x} \p)^2}{2\text{x}(1-\text{x})E}\, z\right)   - \sin\left( \frac{\q\cdot (2 \k -(1-2\text{x}) \q - 2 \text{x} \p)}{2\text{x}(1-\text{x})E} \, z\right)\bigg] \Bigg\}  \, \notag
    \\ & \hspace{4cm}   \times \rho(z) \, [v(\q^2)]^2 \, \left|J(E,\p - \q)\right|^2 \,,
\end{align}
where the color factors can also be verified in the limit of homogeneous matter.

\subsection{Double Born diagrams}

\begin{figure}[t!]
    \centering
    \includegraphics[width=1\textwidth]{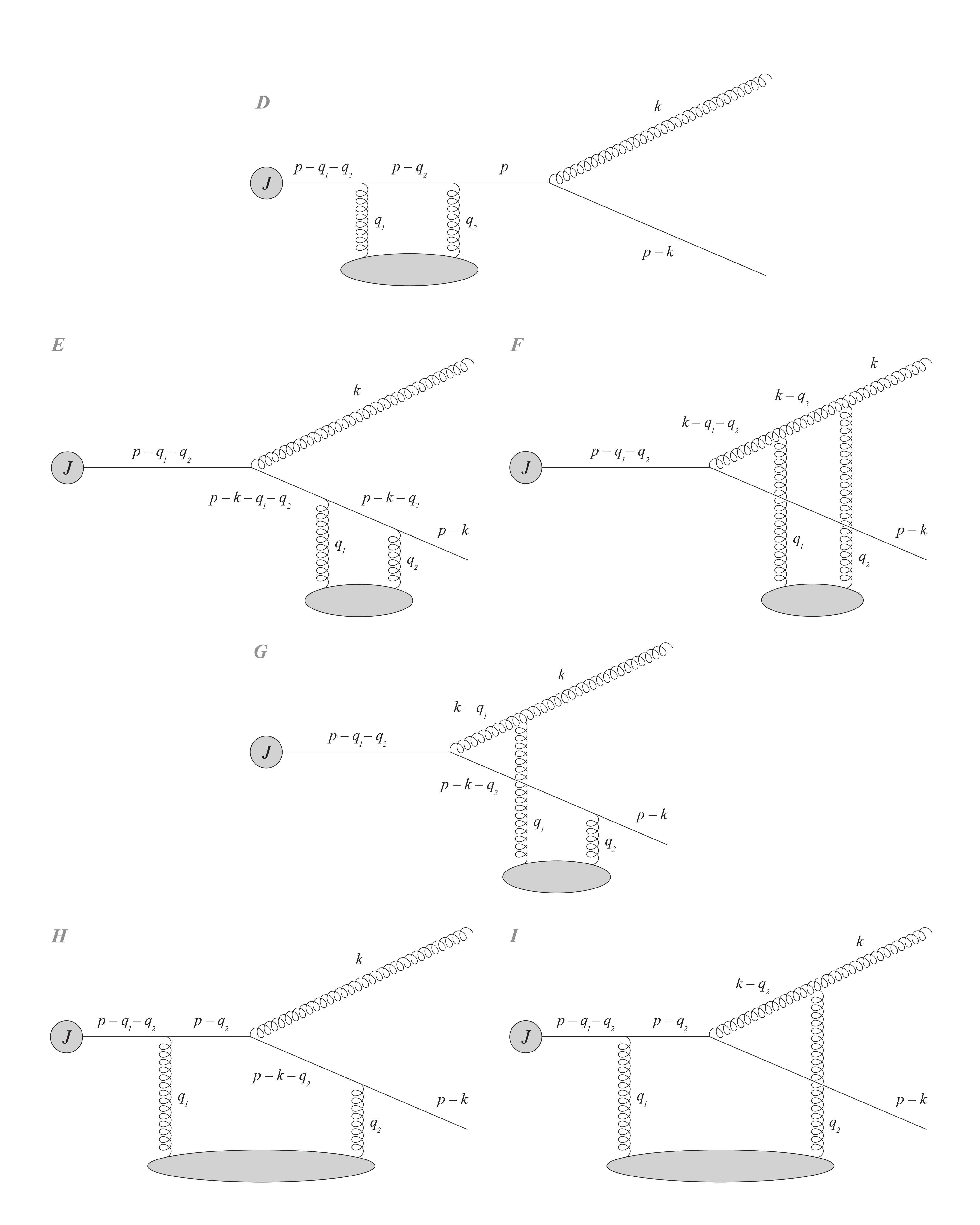}
    \vspace*{-10mm}\caption{The six double-Born diagrams contributing to the medium-induced radiation at the first order in opacity.}
\label{f:matter_DB}
\end{figure}

Here, we consider the six DB contributions in the amplitude of the gluon radiation, see Fig.~\ref{f:matter_DB}. Following \cite{Sadofyev:2021ohn}, we note that after averaging over medium configurations, the propagator poles, governing the $q_{1z}$-integral, cancel $q_{2z}$ in the exponential factors (except for a part of $\cR_G$), and the contour can be closed in both directions. The scattering potentials are needed for convergence, and should be taken into account explicitly. 

%
%

Starting with $\cR_D$, corresponding to the case of two initial-state scatterings, we find its contribution to the squared amplitude:
\begin{align}
\label{Rd initial}
    \langle \cR_D \cR^*_0 \rangle +\text{c.c.} &= i \frac{C^2_F}{N_c} \, g^2 
    \int_{x,\q_1,\q_2} \,  \theta(z) \, \rho(\x,z) \, \frac{4(1-\text{x})^2}{(\k-\text{x}\p)^2} \, J^*(E,\p - \q_1 - \q_2) J(E,\p) \, \notag
    \\ & \hspace{1.5cm} \times \left[e^{-i(\q_1+\q_2)\cdot \x} \, e^{-i\cQ^-_{p-q_1-q_2}\, z} \, \mathcal{I}_D - e^{i(\q_1+\q_2) \cdot \x} \, e^{i\cQ^-_{p-q_1-q_2} \, z} 
    \, \mathcal{I}_D^* \right] \, ,
\end{align}
where the integral $\mathcal{I}_D$ is defined as 
\begin{align}
    \mathcal{I}_D = \int \frac{dq_{2z}}{2\pi}\, \frac{E \, v(q_2) \,v(\tilde{q_1})} {(q_{2z}-\cQ^+_{p-q_2}-i\epsilon)(q_{2z}-\cQ^-_{p-q_2}+i\epsilon)}  
\end{align}
with $ \tilde{q}_{1\mu}=(\u\cdot\q_1, \,\q_1, \,-q_{2z}+\cQ^-_{p-q_1-q_2})$. This integral can be explicitly evaluated, and since it involves no length enhanced terms, we keep only its eikonal part. Then,
\begin{align}\label{ReId&ImID}
    &\im\,\mathcal{I}_D(\q,-\q) \simeq \frac{1}{4}[v(\q^2)]^2\notag\\
    &\re\,\mathcal{I}_D(\q,-\q) \simeq \frac{2(\u\cdot\q)^3-3(\u\cdot\q)(\m^2+\q^2)}{8R_0^3}[v(\q^2)]^2\,,
\end{align}
where $R_0$ is given by \eqref{e:Rdef} at $u_z=0$. One should also note that $\frac{\partial}{\partial (\q_1+\q_2)}\,\re\, \mathcal{I}_{D}\Big|_{\q_1=-\q_2}=0$, and it results in no eikonal gradient corrections. Thus, keeping only the terms with non-zero angular average, we find
\begin{align}
     \langle \cR_D \cR^*_0\rangle + \text{c.c.} &= - \frac{C^2_F}{ N_c} \, g^2 \int_0^L dz \int_\q \,  \frac{2(1-\text{x})^2}{(\k-\text{x}\p)^2} \left[1 - \hat{\g} \cdot \left(\u - \frac{\p}{E}\right) \, z\right] \, \notag
    \\ & \hspace{1.5cm} \times \cos\left(\frac{(\k-\text{x}\p)^2}{2\text{x}(1-\text{x})E} \, z\right) \rho(z) \, [v(\q^2)]^2 \left|J(E,\p)\right|^2 \, .
\end{align}

Similarly, for the contribution to the amplitude squared with two final-state scatterings in the quark line $\cR_E$, we find\footnote{Taking the first $q_{1z}$-integral, we pick up two residues, corresponding to the poles $\cQ^-_{p-q_1-q_2}$ and $\cQ^-_{p-k-q_1-q_2}$.  At eikonal order, the scattering potentials look the same for both residues and the Fourier phases can be pulled out as a common factor for $\mathcal{I}_E$.}
\begin{align}
    \left\langle \cR_E \cR^*_0\right\rangle+\text{c.c.} &= -i \frac{C^2_F}{N_c} \, g^2 \int_{x,\q_1,\q_2}  \theta(z) \, \rho(\x,z) \, \notag
    \\ & \hspace{-1cm} \times \frac{4(1-\text{x})^2\, (\k-\text{x}\p)\cdot(\k-\text{x}(\p-\q_1-\q_2))}{(\k-\text{x}\p)^2 (\k-\text{x}(\p-\q_1-\q_2))^2} \, J(E,\p) J(E,\p - \q_1 - \q_2) \, \notag
    \\ & \hspace{0cm} \times \Bigg[e^{-i(\q_1+\q_2) \cdot \x} \left( e^{-i\cQ^-_{p-q_1-q_2} \, z} - e^{-i\cQ^-_{p-k-q_1-q_2} \, z}\right) \mathcal{I}_E \, \notag
    \\ & \hspace{4cm} - e^{i(\q_1+\q_2) \cdot \x} \left(e^{i\cQ^-_{p-q_1-q_2} \, z} - e^{i\cQ^-_{p-k-q_1-q_2} \, z}\right) \, \mathcal{I}_E^*\, \Bigg] \, ,
\end{align}
where
\begin{align}
\label{I_E}
    \mathcal{I}_E = \int \frac{dq_{2z}}{2\pi}\,\frac{(1-\text{x})E \, v(q_2) \, v(\tilde{q}_1)}{(q_{2z} - \cQ^+_{p-k-q_2}-i\epsilon)(q_{2z} - \cQ^-_{p-k-q_2}+i\epsilon)} \, 
\end{align}
with $\tilde{q}_{1\mu} = (\u \cdot \q_1, \,\q_1, \,-q_{2z} + \cQ^-_{p-k-q_1-q_2} )$. Evaluating this integral, one may note that at the leading order in the eikonal expansion $\mathcal{I}_E\simeq\mathcal{I}_D$, and we can again utilize \eqref{ReId&ImID}. Thus, expanding in gradients, we find that
\begin{align}
    \langle\cR_E \cR^*_0\rangle + \text{c.c.} &= - \frac{C^2_F}{ N_c} \, g^2 \int_0^L dz \int_\q \,  \frac{2(1-\text{x})^2}{(\k-\text{x}\p)^2} \, \Bigg\{ \left[1 - \hat{\g} \cdot \u \, z\right] \, \notag
    \\ & \hspace{-2cm} \times \left[1-\cos\left(\frac{(\k-\text{x}\p)^2}{2\text{x}(1-\text{x})E}\,z\right) \right] + \hat{\g} \cdot \left( \frac{\p-\k}{(1-\text{x})E} z - \frac{\p}{E} z \,\cos\left(\frac{(\k-\text{x}\p)^2}{2\text{x}(1-\text{x})E}\,z\right)\right) \, \notag
    \\ & \hspace{-1cm} + \text{x} \, \hat{\g} \cdot \frac{\k-\text{x}\p}{(\k-\text{x}\p)^2} \, \sin\left(\frac{(\k-\text{x}\p)^2}{2\text{x}(1-\text{x})E} \, z\right) \Bigg\} \, \rho(z) \, [v(\q^2)]^2 \, \left|J(E,\p)\right|^2 \,.
\end{align}

Turning to the contribution with two final-state scatterings in the gluon line, we have to deal with a product of two gluon propagators and two three-gluon vertices. As in the case of $\cR_C$ above, this product does not depend on the $z$-component of momenta since $N^{\mu\nu}$, $u_\mu$, and $\epsilon^*_\mu$ are transverse to $n_\mu$, resulting in no new poles. Averaging the corresponding contribution to the amplitude squared, we write it as
\begin{align}
   &\langle \cR_F \cR^*_0 \rangle + \text{c.c.} =-i \frac{C_F}{2} \, g^2 \, \int_{x,\q_1,\q_2} \theta(z) \, \rho(\x,z) \,  \notag
    \\ & \hspace{1cm} \times \frac{8 (1-\text{x})^2 \, (\k-\text{x}\p) \cdot (\k -(1-\text{x})(\q_1 + \q_2)-\text{x}\p)}{(\k-\text{x}\p)^2 (\k -(1-\text{x})(\q_1 + \q_2)-\text{x}\p)^2} \, J(E,\p) J^*(E,\p-\q_1-\q_2) \, \notag
    \\ & \hspace{2cm} \times \bigg[ e^{-i(\q_1+\q_2) \cdot \x} \left(e^{-i\cQ^-_{p-q_1-q_2} \, z} - e^{-i\cQ^-_{k-q_1-q_2} \, z}\right) \mathcal{I}_F\notag
    \\ & \hspace{6cm} - e^{i(\q_1+\q_2) \cdot \x} \left(e^{i\cQ^-_{p-q_1-q_2} \, z} - e^{i\cQ^-_{k-q_1-q_2} \, z}\right) \mathcal{I}_F^* \bigg] \, ,
\end{align}
where
\begin{align}
    \mathcal{I}_F = \int \frac{dq_{2z}}{2\pi} \,\frac{\text{x}E \, v(q_2) \, v(\tilde{q}_1) }{(q_{2z}-\cQ^+_{k-q_2}-i\epsilon)(q_{2z}-\cQ^-_{k-q_2}+i\epsilon)}  \, ,
\end{align}
with $\tilde{q}_{1\mu} = (\u \cdot \q_1, \,\q_1, \,-q_{2z} + \cQ^-_{k-q_1-q_2})$. At the leading order in the eikonal expansion, one finds that $\mathcal{I}_F\simeq\mathcal{I}_D$, and we can again utilize \eqref{ReId&ImID}. Expanding in gradients and treating $\x$-integral in the same fashion, one finds
\begin{align}
    &\langle \cR_F \cR^*_0 \rangle + \text{c.c.} = - C_F \, g^2 \int_0^L dz \int_\q \,  \frac{2(1-\text{x})^2}{(\k-\text{x}\p)^2} \Bigg\{ \left[1 - \hat{\g} \cdot \u \, z\right] \, \notag
    \\ & \hspace{1cm} \times \left[1 - \cos\left(\frac{(\k-\text{x}\p)^2}{2\text{x}(1-\text{x})E} \, z\right) \right] +\hat{\g} \cdot \left(\frac{\k}{\text{x}E} z -\frac{\p}{E} z \, \cos\left(\frac{(\k-\text{x}\p)^2}{2\text{x}(1-\text{x})E} \, z\right)\right) \, \notag
    \\ & \hspace{2cm} - (1-\text{x}) \, \hat{\g} \cdot  \frac{\k-\text{x}\p}{(\k-\text{x}\p)^2} \, \sin\left(\frac{(\k-\text{x}\p)^2}{2\text{x}(1-\text{x})E} \, z\right) \Bigg\} \, \rho(z) \, [v(\q^2)]^2 \, \left|J(E,\p)\right|^2 \, .
\end{align}

Now we can turn to a more involved diagram, having two final-state scatterings in different lines. The corresponding contribution to the amplitude squared can be written as 
\begin{align}
    &\langle \cR_G \cR^*_0 \rangle + \text{c.c.} = i \frac{C_F}{2} \, g^2 \, \, \int_{x,\q_1,\q_2} \theta(z) \, \rho(\x,z) \, \frac{4\,(1-\text{x}) \, (\k - \text{x}\p) \cdot (\k - \q_2 - \text{x}(\p-\q_1-\q_2))}{(\k-\text{x}\p)^2} \,  \notag
    \\ & \hspace{2cm} \times J(E,\p) J^*(E,\p - \q_1 - \q_2) \, \bigg[ e^{-i(\mathbf{q}_1+\mathbf{q}_2)\cdot\mathbf{x}} 
    \, \left(e^{-i (\cQ^-_{k-q_2}+\cQ^-_{p-k-q_1})\,z} - e^{-i\cQ^-_{p-q_1-q_2}\,z}\right) \mathcal{I}_G \, \notag
    \\ & \hspace{4cm} - e^{i(\mathbf{q}_1+\mathbf{q_2})\cdot\mathbf{x}} 
    \, \left(e^{i (\cQ^-_{k-q_2}+\cQ^-_{p-k-q_1})\,z} - e^{i\cQ^-_{p-q_1-q_2}\,z}\right) \mathcal{I}_G^*   \bigg] \, ,
\end{align}
where $\mathcal{I}_G\equiv \left(e^{-i (\cQ^-_{k-q_2}+\cQ^-_{p-k-q_1}) \, z} - e^{-i\cQ^-_{p-q_1-q_2} \, z}\right)^{-1} \mathcal{\bar{I}}_G$, and
\begin{align}
    \mathcal{\bar{I}}_G &\equiv \int \frac{dq_{2z}}{2\pi} \,\frac{E \, v(q_2)}{(q_{2z}-\cQ^+_{k-q_2}-i\epsilon)(q_{2z}-\cQ^-_{k-q_2}+i\epsilon)(q_{2z}-\cQ^+_{p-q_1-q_2}+\cQ^-_{p-k-q_1}-i\epsilon)} \, \notag
    \\ & \hspace{1cm} \Bigg[ \frac{v(q'_1) \, e^{-i(\cQ^+_{p-k-q_1}+q_{2z}) \, z}}{ q_{2z} + \cQ^-_{p-k-q_1} - \cQ^+_{p-q_1-q_2}-i\epsilon} - \frac{(1-x) \, v(\tilde{q}_1) \, e^{-i \cQ^+_{p-q_1-q_2} \, z}}{-q_{2z} + \cQ^+_{p-q_1-q_2} - \cQ^+_{p-k-q_1}-i\epsilon} \Bigg]
\end{align}
with $q'_{1\mu} = (\u \cdot \q_1, \,\q_1, \cQ^+_{p-k-q_1})$ and $\tilde{q}_{1\mu} = (\u \cdot \q_1, \,\q_1, \,-q_{2z} + \cQ^+_{p-q_1-q_2})$. Evaluating this integral, we close the contour below the real axis, so the explicit exponential factor can be utilized. In the eikonal limit, we further find that
\begin{align}
    &\im\,\mathcal{I}_G(\q_1,\q_2) \simeq -\frac{1}{2}\frac{(1-\text{x})}{(\k-\q_2-\text{x}(\p-\q_1-\q_2))^2}v(\q_1^2)v(\q_2^2)\notag\\
    &\re\,\mathcal{I}_G(\q_1,\q_2) \simeq 0\,.
\end{align}
Thus, the corresponding contribution to the amplitude squared reads
\begin{align}
    &\hspace{-0.75cm}\langle\cR_G \cR^*_0\rangle + \text{c.c.} = C_F \, g^2 \int_0^L dz \int_\q \,  \frac{2(1-\text{x})^2 \, (\k - \text{x}\p) \cdot (\k + \q -\text{x}\p)}{(\k-\text{x}\p)^2 (\k + \q -\text{x}\p)^2} \, \notag
    \\ & \hspace{-0.5cm} \times \Bigg\{ \left[1 - \hat{\g} \cdot \u \, z\right] \left[\cos\left(\frac{\q \cdot (2(\k-\text{x}\p) + \q)}{2\text{x}(1-\text{x})E} \, z\right) - \cos\left(\frac{(\k-\text{x}\p)^2}{2\text{x}(1-\text{x})E} \, z\right) \right]  \, \notag
    \\ & \hspace{-0.25cm} + \hat{\g} \cdot \left(\frac{\p-\k-\q}{2(1-\text{x})E} z + \frac{\k+\q}{2\text{x}E} z\right)  \cos\left(\frac{\q \cdot (2(\k-\text{x}\p) + \q)}{2\text{x}(1-\text{x})E} \, z\right)  - \hat{\g} \cdot \frac{\p}{E}z \, \cos\left(\frac{(\k-\text{x}\p)^2}{2\text{x}(1-\text{x})E} \, z\right) \,\notag
    \\ & \hspace{0cm} +\left(\text{x}-\frac{1}{2}\right)  \, \hat{\g} \cdot \left( 2\frac{\k+\q-\text{x}\p}{(\k + \q -\text{x}\p)^2} -\frac{\k-\text{x}\p}{(\k-\text{x}\p) \cdot (\k+\q-\text{x}\p)} \right) \, \notag
    \\ & \hspace{0.25cm} \times \bigg[\sin\left(\frac{\q \cdot (2 (\k-\text{x}\p) + \q)}{2\text{x}(1-\text{x})E} \, z\right) + \sin\left(\frac{(\k-\text{x}\p)^2}{2\text{x}(1-\text{x})E} \, z\right) \bigg] \Bigg\} \, \rho(z) \, [v(\q^2)]^2 \, \left|J(E,\p)\right|^2 \, ,
\end{align}
where the gradient corrections have been treated in the same way as before.

The two remaining DB diagrams are not expected to contribute in the eikonal limit, since they have the gluon emission vertex in between the two interactions, attached to the same source. Starting with the case of initial- and final-state scatterings in the quark lines, we can write the corresponding contribution as
\begin{align}
     &\hspace{-0.5cm}\langle \cR_H \cR^*_0\rangle + \text{c.c.} = i \frac{C_F}{N_c^2} \, g^2 \,  \int_{x,\q_1,\q_2}  \theta(z) \, \rho(\x,z)  \, \frac{4(1-\text{x})^2 \,(\k-\text{x}\p)\cdot (\k-\text{x}(\p-\q_2))}{\text{x} (\k-\text{x}\p)^2} \notag
    \\ & \hspace{0cm} \times \left[ e^{-i(\q_1+\q_2) \cdot \x} \, e^{-i \cQ^-_{p-q_1-q_2} z} \, \mathcal{I}_H - e^{i(\q_1+\q_2) \cdot \x} e^{i \cQ^-_{p-q_1-q_2} z} \, \mathcal{I}_H^* \right] \, J(E,\p) J^*(E,\p -\q_1 -\q_2) ,
\end{align}
where
\begin{align}
    \mathcal{I}_H = \int\frac{dq_{2z}}{2\pi} \, \frac{E \, v(\tilde{q}_1) \, v(q_2)}{(q_{2z} - \cQ^+_{p-q_2}-i\epsilon)(q_{2z} - \cQ^-_{p-q_2}+i\epsilon) \, (q_{2z} - \cQ^+_{p-k-q_2}-i\epsilon)(q_{2z} - \cQ^-_{p-k-q_2}+i\epsilon)} \, ,
\end{align}
and $\tilde{q}_{1\mu}=(\u\cdot\q_1,\, \q_1,\, -q_{2z} + \cQ^-_{p-q_1-q_2})$. Evaluating this integral, we indeed find that it is energy suppressed $\mathcal{I}_H=\mathcal{O}\left(\frac{\perp}{E}\right)$, and $\cR_H$ does not contribute in the considered limit.

Similarly, turning to the last contribution with initial-state scattering in the quark line and final-state scattering in the gluon line, we write it as
\begin{align}
   \langle \cR_I \cR^*_0\rangle  + \text{c.c.} &= - i \frac{C_F}{2} \, g^2 \,  \int_{x,\q_1,\q_2} \theta(z) \, \rho(\x,z) \, \frac{4(1-\text{x}) \, (\k - \text{x}\p) \cdot (\k-(1-\text{x})\q_2-\text{x} \p)}{(\k-\text{x}\p)^2} \notag
    \\ & \hspace{-2.2cm} \times \left[e^{-i(\q_1+\q_2) \cdot \x} \, e^{-i \cQ^-_{p-q_1-q_2} z} \, \mathcal{I}_I\, - e^{i(\q_1+\q_2) \cdot \x} \, e^{i \cQ^-_{p-q_1-q_2} z} \, \mathcal{I}_I^*\,\right] \, J(E,\p) J^*(E,\p-\q_1-\q_2) \, ,
\end{align}
where
\begin{align}
     \mathcal{I}_I = \int\frac{dq_{2z}}{2\pi} \, \frac{E \, v(\tilde{q}_1) \, v(q_2)}{(q_{2z} - \cQ^+_{p-q_2}-i\epsilon)(q_{2z} - \cQ^-_{p-q_2}+i\epsilon) \, (q_{2z} - \cQ^+_{k-q_2}-i\epsilon)(q_{2z} - \cQ^-_{k-q_2}+i\epsilon)} \, ,
\end{align}
and $\tilde{q}_{1\mu}=(\u\cdot\q_1,\, \q_1,\, -q_{2z} + \cQ^-_{p-q_1-q_2})$. One may readily show that this integral is also energy suppressed, and cannot contribute to the squared amplitude in the considered limit.

\subsection{Final distribution and its properties} \label{SubSecFinalDistribRad}

The final-state parton distribution can now be expressed through the emission amplitude squared as
\begin{align}
\label{dNforRad}
    E\, \frac{d\cN^{(1)}}{d^2k \, d\text{x} \, d^2p \, dE} \equiv \frac{1}{\left[2(2\pi)^3\right]^2} \, \frac{1}{\text{x}(1-\text{x})} \, \langle|\cR_{N=1}|^2\rangle \, ,
\end{align}
where the superscript indicates that only $N=1$ terms are included. It depends on the source of energetic quarks, and allows for the study of how an ensemble of quarks radiates while propagating through the matter. In this work, we assume that the initial distribution $E\, \frac{d\cN^{(0)}}{d^2p \, dE}$ is a slowly varying function of the transverse momentum, setting $E\, \frac{d\cN^{(0)}}{d^2(p-q) \, dE}\simeq E\, \frac{d\cN^{(0)}}{d^2p \, dE}$.  

In what follows, we use a set of shorthand notation in order to make the final expressions more compact. First, we introduce the characteristic LPM phases, entering the final-state distribution: 
\begin{align}
&\phi=\frac{(\k -\text{x}(\p-\q))^2}{2 \text{x} (1-\text{x}) E} \, z\,,&  &\bar{\phi}=\frac{(\k - (1-\text{x})\q-\text{x}\p)^2}{2 \text{x} (1-\text{x}) E} \, z \,, &\notag\\
&\phi_0=\frac{(\k -\text{x}\p)^2}{2 \text{x} (1-\text{x}) E} \, z\,,& &\phi_G=\frac{\q\cdot(2(\k-\text{x}\p)+\q)}{2\text{x}(1-\text{x})E}z\,.&\notag
\end{align}
It is also convenient to re-express the light-front wave functions through their normalized arguments:
\begin{align}
&\boldsymbol{\kappa}=\frac{\k-\text{x}(\p-\q)}{(\k-\text{x}(\p-\q))^2}\,,& &\bar{\boldsymbol{\kappa}}=\frac{\k-(1-\text{x})\q-\text{x}\p}{(\k-(1-\text{x})\q-x\p)^2}\,,&\notag\\
&\boldsymbol{\kappa}_0=\frac{\k-\text{x}\p}{(\k-\text{x}\p)^2}\,,& &\boldsymbol{\kappa}_G=\frac{\k + \q -\text{x}\p}{(\k + \q -\text{x}\p)^2}\,.&\notag
\end{align}
Finally, we associate the particular momentum structures in the gradient corrections with the diagrams where they appear for the first time
\begin{align}
&~~~~\tvec{D}_A=\frac{\p-\q}{E}z\,,~~~~~~~~ \tvec{D}_B=\frac{\p-\q}{2E}z+\frac{\p-\k-\q}{2(1-\text{x})E}z\,,~~~~~~~~\tvec{D}_C=\frac{\p-\q}{2E}z+\frac{\k-\q}{2\text{x} E}z\,,&\notag\\
&\tvec{D}_D=\frac{\p}{E}z\,,~~~~~~~~\tvec{D}_E=\frac{\p-\k}{(1-\text{x})E}z\,,~~~~~~~~\tvec{D}_F=\frac{\k}{\text{x} E}z\,,~~~~~~~~\tvec{D}_G=\frac{\p-\k-\q}{2(1-\text{x})E}z  + \frac{\k+\q}{2\text{x}E}z\,.&\notag
\end{align}

Let us now focus on the limit of static inhomogeneous matter, setting $\u=0$ and keeping the leading subeikonal (but length enhanced) terms. The resulting distribution extends the consideration in \cite{Sadofyev:2021ohn}, where the gradient corrections to the jet broadening were studied within the opacity expansion, to the case of in-medium branching. It is obtained in the full kinematics, although only up to the first order in opacity, extending the recent results for the all-order soft gluon spectrum in inhomogeneous matter in \cite{Barata:2023qds}. Combining the contributions to the amplitude squared derived in Sec.~\ref{SecIII:Branching} at $\u=0$, we find
\begin{align} \label{e:dN_u=0}
     \hspace{-0.5cm}E\, \frac{d\cN^{(1)}}{d^2k \, d\text{x} \, d^2p \, dE} & = \frac{(1-\text{x})g^2}{(2\pi)^3\,\text{x}}\frac{C_F^2}{N_c}\left(E\, \frac{d\cN^{(0)}}{d^2p \, dE}\right)\int_0^L dz \, \int_{\q} \, \notag
    \\ & \hspace{-2.75cm} \times\Bigg\{\boldsymbol{\kappa}_0^2\left(1+\hat{\g}\cdot\tvec{D}_A\right)  +  2 \boldsymbol{\kappa}^2\,\left(1 -\cos\phi \right) \left(1+  \hat{\g} \cdot \tvec{D}_B\right) + 2\frac{N_c}{C_F}\bar{\boldsymbol{\kappa}}^2   \, \left(1 -\cos\bar{\phi} \right) \left(1+\hat{\g} \cdot \tvec{D}_C\right) \, \notag
    \\ & \hspace{-2cm} +\frac{\boldsymbol{\kappa}_0\cdot\boldsymbol{\kappa}}{2C_FN_c}  \left[ 2\left(1 - \cos\phi\right)  +  2 \, \hat{\g} \cdot \left(\tvec{D}_A-\tvec{D}_B \cos\phi\right) - \text{x} \, \hat{\g}\cdot\left(2\boldsymbol{\kappa}-\boldsymbol{\kappa}^2\frac{\boldsymbol{\kappa}_0}{\boldsymbol{\kappa}_0\cdot\boldsymbol{\kappa}}\right)\sin\phi \right] \notag
    \\ & \hspace{-2cm} - \frac{N_c}{2C_F}\boldsymbol{\kappa}_0\cdot\bar{\boldsymbol{\kappa}} \, \left[ 2\left(1 -\cos\bar{\phi} \right)  +2 \, \hat{\g} \cdot \left(\tvec{D}_A - \tvec{D}_C \cos\bar{\phi}\right)  +(1-\text{x})\,\hat{\g}\cdot\left(2\bar{\boldsymbol{\kappa}}-\bar{\boldsymbol{\kappa}}^2\frac{\boldsymbol{\kappa}_0}{\boldsymbol{\kappa}_0\cdot\bar{\boldsymbol{\kappa}}}\right)\sin\bar{\phi}\right] \notag
    \\ &\hspace{-2cm} - \frac{N_c}{2C_F}\boldsymbol{\kappa}\cdot\bar{\boldsymbol{\kappa}} \Bigg[2\left(1 + \cos\left(\phi-\bar{\phi}\right) - \cos\bar{\phi} - \cos\phi\right) \, \notag
    \\ &\hspace{0.5cm} + 2 \, \hat{\g} \cdot \left(\tvec{D}_A  - \tvec{D}_B \cos\phi - \tvec{D}_C \cos\bar{\phi} + \left(\tvec{D}_B+\tvec{D}_C-\tvec{D}_A\right) \cos\left(\phi-\bar{\phi}\right)  \right)\notag
    \\ &\hspace{0.5cm} - \hat{\g}\cdot\bigg(2(1-\text{x})\bar{\boldsymbol{\kappa}}-\frac{\boldsymbol{\kappa}_0}{\boldsymbol{\kappa}_0^2}\frac{\bar{\boldsymbol{\kappa}}^2\boldsymbol{\kappa}^2}{\boldsymbol{\kappa}\cdot\bar{\boldsymbol{\kappa}}}+2\text{x}\boldsymbol{\kappa}\bigg) \bigg(\sin\phi -\sin\bar{\phi}-\sin\left(\phi-\bar{\phi}\right)\bigg)\Bigg]  \, \notag
    \\ & \hspace{-2cm} - \boldsymbol{\kappa}_0^2 \, \cos\phi_0 \left(1 + \hat{\g} \cdot \tvec{D}_D\right)  - \boldsymbol{\kappa}_0^2 \, \left[ \left( 1 - \cos\phi_0\right) + \hat{\g} \cdot \left( \tvec{D}_E -\tvec{D}_D\cos\phi_0\right) + \text{x}\,\hat{\g} \cdot\boldsymbol{\kappa}_0\,\sin\phi_0\right]\notag
    \\ & \hspace{-2cm} - \frac{N_c}{C_F}\boldsymbol{\kappa}_0^2 \, \left[ \left( 1 - \cos\phi_0\right) + 
    \hat{\g} \cdot \left(\tvec{D}_F -\tvec{D}_D\cos\phi_0\right) -(1-\text{x})\,\hat{\g} \cdot\boldsymbol{\kappa}_0\,\sin\phi_0\right]\notag
    \\ & \hspace{-2cm} +  \frac{N_c}{C_F}\boldsymbol{\kappa}_0\cdot\boldsymbol{\kappa}_G \bigg[\left(\cos\phi_G - \cos\phi_0\right)  +  \hat{\g} \cdot \left(  \tvec{D}_G\cos\phi_G  - \tvec{D}_D\cos\phi_0\right)\notag
    \\ & \hspace{0cm} +\left(\text{x}-\frac{1}{2}\right)\,\hat{\g}\cdot \bigg(2\,\boldsymbol{\kappa}_G-\boldsymbol{\kappa}_G^2\frac{\boldsymbol{\kappa}_0}{\boldsymbol{\kappa}_0\cdot\boldsymbol{\kappa}_G}\bigg) \left(\sin\phi_G+\sin\phi_0\right) \bigg] \Bigg\}\, \rho(z) \, [v(\q^2)]^2 \, .
\end{align}
Taking the limit of homogeneous matter with $\hat \g=0$, we can readily check that this expression agrees with the result for the $N=1$ in-medium branching, see e.g. \cite{Sievert:2018imd}. 

To make the features of the final-state parton distribution more apparent, it is instructive to consider the small-x limit of \eqref{e:dN_u=0}, where the distribution is known to take a particularly simple form in the case of homogeneous matter \cite{Gyulassy:2000er}. In this limit $\phi=\phi_0$, $\phi_G=\bar{\phi}|_{\q\to-\q}-\phi_0$, $\tvec{\kappa}=\tvec{\kappa}_0$, and $\tvec{\kappa}_G=\bar{\tvec{\kappa}}|_{\q\to-\q}$. Keeping only the subeikonal terms which scale as $\frac{1}{\text{x}E}$, we set $\tvec{D}_A=\tvec{D}_B=\tvec{D}_D=\tvec{D}_E=0$, then
\begin{align}\label{e:dN_small-x_u=0}
    \hspace{-0.25cm}E\, \frac{d\cN^{(1)}}{d^2k \, d\text{x} \, d^2p \, dE} & = \frac{g^2\,C_F}{(2\pi)^3\,\text{x}}\left(E\, \frac{d\cN^{(0)}}{d^2p \, dE}\right)\int_0^L dz \, \int_{\q}  \notag
    \\ & \hspace{-2.75cm} \times\Bigg\{\frac{2\,\k\cdot\q}{\k^2(\k-\q)^2} \, \left(1 -\cos\left(\frac{(\k-\q)^2}{2\text{x}E}z\right) \right)\left(1+\frac{\hat{\g} \cdot (\k-\q)}{2\text{x}E}z\right)- \frac{\hat{\g}\cdot\k}{\k^2} \, \left[
    \frac{z}{\text{x}E} -\frac{1}{\k^2}\sin\left(\frac{\k^2}{2\text{x}E}z\right)\right]\notag
    \\ & \hspace{-2.5cm}  + \frac{\k\cdot(\k-\q)}{\k^2(\k-\q)^2}\, \left[\frac{\hat{\g}\cdot(\k-\q)}{\text{x}E}z - \hat{\g}\cdot\left(2\frac{\k-\q}{(\k-\q)^2}-\frac{\k}{\k\cdot(\k-\q)}\right)\sin\left(\frac{(\k-\q)^2}{2\text{x}E}z\right)\right]\Bigg\} \, \rho(z) \, [v(\q^2)]^2 \,.
\end{align}
This expression can be compared with the results of \cite{Barata:2023qds}, and, after some algebra, one can show that \eqref{e:dN_small-x_u=0} precisely agrees with the $N=1$ part of the small-x resummed parton distribution.

Turning to the flow-gradient effects, we take the eikonal limit, and note that only the same multiplicative factor in the integrand of the amplitude squared survives. Then, the final-state distribution reads
\begin{align}
\label{dNgu}
    E\, \frac{d\cN^{(1)}}{d^2k \, d\text{x} \, d^2p \, dE} & = \frac{(1-\text{x}) g^2}{(2\pi)^3\,\text{x}}\frac{C_F^2}{N_c}\left(E\, \frac{d\cN^{(0)}}{d^2p \, dE}\right)\int_0^L dz \, \int_{\q} \, \left(1-\hat{\g}\cdot\u\,z\right)\notag
    \\ & \hspace{-2.5cm} \times\Bigg\{\boldsymbol{\kappa}_0^2  +  2 \boldsymbol{\kappa}^2\,\left(1 -\cos\phi \right) + 2\frac{N_c}{C_F}\bar{\boldsymbol{\kappa}}^2   \, \left(1 -\cos\bar{\phi} \right)  +\frac{\boldsymbol{\kappa}_0\cdot\boldsymbol{\kappa}}{C_FN_c}\,  \left(1 - \cos\phi\right) - \frac{N_c}{C_F}\boldsymbol{\kappa}_0\cdot\bar{\boldsymbol{\kappa}} \, \left(1 -\cos\bar{\phi} \right)\notag
    \\ & \hspace{-1.25cm}  - \frac{N_c}{C_F}\boldsymbol{\kappa}\cdot\bar{\boldsymbol{\kappa}} \,\left(1 + \cos\left(\phi-\bar{\phi}\right) - \cos\bar{\phi} - \cos\phi\right) - \boldsymbol{\kappa}_0^2 \, \cos\phi_0  - \boldsymbol{\kappa}_0^2 \, \left( 1 - \cos\phi_0\right) \notag
    \\ & \hspace{0cm} - \frac{N_c}{C_F}\boldsymbol{\kappa}_0^2 \,\left( 1 - \cos\phi_0\right) +  \frac{N_c}{C_F}\boldsymbol{\kappa}_0\cdot\boldsymbol{\kappa}_G \,\left(\cos\phi_G - \cos\phi_0\right)\Bigg\}\, \rho(z) \, [v(\q^2)]^2 .
\end{align}
One readily observes that this modification of the distribution results in a multiplicative modification of the radiation rate due to the jet-medium interactions, and, consequently, in a modification of the induced radiative energy loss, {\it c.f.} \cite{Gyulassy:2000er}.

%
\begin{figure}[t!]
    \centering
    \includegraphics[width=1\textwidth]{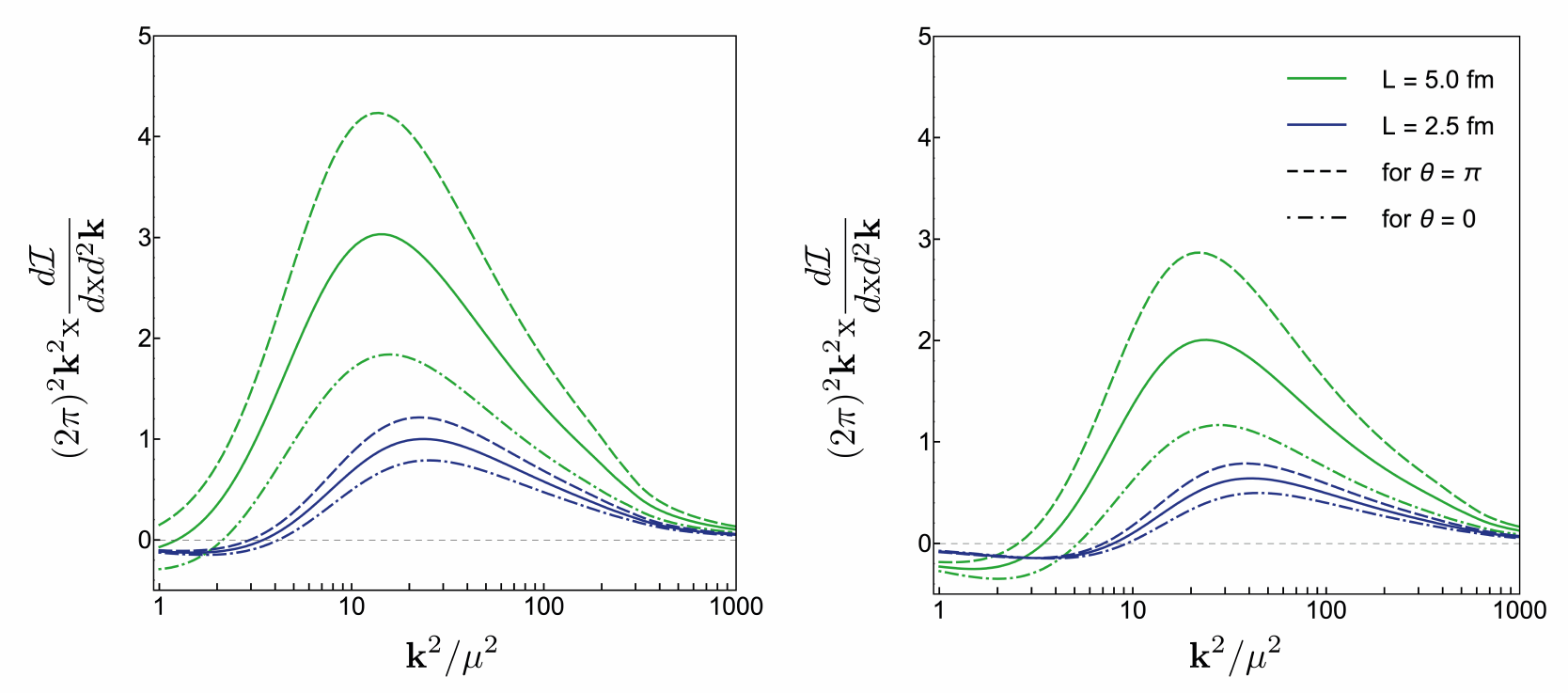}
    \vspace*{-10mm}\caption{The (rescaled) medium-induced soft gluon spectrum is plotted for two energies $E=50\,$GeV (left) and $E=100\,$GeV (right). The colors distinguish the medium length, while the mean free path is kept fixed ($\chi=3$ at $L=5\,$fm). The solid lines correspond to the homogeneous (or no transverse flow) limit, while the dashed lines correspond to $\u$ and $\boldsymbol{\na}T$ being parallel or antiparallel.}
\label{f:GLVnew}
\end{figure}
%

Estimating the effect of the mixed term in the spectrum, we focus on the small-x limit. Then, the final state distribution can be factorized into the initial quark distribution and an emission spectrum $d\mathcal{I}^{(1)}$, defined by
\begin{align}
    \text{x} E \,  \frac{d\cN^{(1)}}{d^2k \, d\text{x} \, d^2p \, dE} \equiv \text{x} \, \frac{d\mathcal{I}^{(1)}}{d\text{x}d^2k} \, E\, \frac{d\cN^{(0)}}{d^2p \, dE} \, .
\end{align}
Following \cite{Gyulassy:2000fs,Gyulassy:2000er} we choose a smooth longitudinal profile for the source density $\rho(\x,z)= 2 \rho_0(\x) e^{-\frac{2z}{L}}$, treating the $z$-integral analitically. The resulting medium induced gluon spectrum reads
\begin{align} \label{e:dIdxd2k}
    \text{x} \, \frac{d\mathcal{I}^{(1)}}{d\text{x}d^2k} & = \frac{4\a_s\,\chi\,N_c}{\pi} \int_{\q} \,\frac{2\,\k\cdot\q}{\k^2(\q^2+\m^2)^2} \,\frac{L^3(\k-\q)^2}{L^2(\k-\q)^4 + 16 \text{x}^2 E^2}\notag
    \\ & \times  \left[1+\left(\frac{L^2(\k-\q)^4}{L^2(\k-\q)^4 + 16 \text{x}^2 E^2}-\frac{3}{2}\right) L \, (\g\cdot\u)\right]\,,
\end{align}
where we have introduced opacity $\chi = \frac{C_F g^4 \rho_0}{2N_c 4 \pi \mu^2}L$, and replaced $\hat{\g}$ by $\g=\frac{\boldsymbol{\na}T}{T}\left(3-\frac{4}{(\q^2+\m^2)}\right)$, assuming that $\rho_0\sim T^3$ and $\mu\simeq g T$, neglecting gradients of the transverse flow $\u$. 

In Fig.~\ref{f:GLVnew}, we plot the spectrum \eqref{e:dIdxd2k} for two energies $E=50\,$GeV (left) and $E=100\,$GeV (right), while for each energy we also show two different medium length $L=5\,$fm and $L=2.5\,$fm, keeping the mean free path $\lambda=\frac{L}{\chi}$ fixed. We set $\a_s=0.3$, $\mu=0.6\,$GeV, $\text{x}=0.1$, and assume that $\chi=3$ at $L=5\,$fm. For a qualitative estimate of the mixed flow-gradient term, we take $|\u|=0.3$, $\frac{|\boldsymbol{\na}T|}{T^2}=0.1$, and $T=0.3\,$GeV. One may readily see that the modification of the spectrum could be substantial even for moderate flows and anisotropies, especially for larger systems.

\section{Discussion and Conclusions}
In this work, we have studied the mutual effect of the transverse flow and matter gradients on jet momentum broadening and in-medium branching processes. We have derived the momentum broadening distribution up to first order in gradients, including the gradients of the longitudinal and transverse flow velocities, and keeping the leading subeikonal corrections. We have also evaluated the leading gradient corrections to the medium-induced gluon spectrum in the full kinematics. These results are obtained within the opacity expansion framework, following the logic of the formalism developed in \cite{Sadofyev:2021ohn}, and extending it.

As we have shown, the interplay of the flow and gradient effects results in the leading modification of the final parton distributions and their even moments. For instance, the jet quenching parameter $\hat q$ is rescaled by an overall factor \eqref{e:qhat_modified}, which may substantially modify its value in the homogeneous static limit. Indeed, let us focus on the contribution proportional to the gradient of the source density $\boldsymbol{\na}\rho$. Assuming that $\rho\sim T^3$, we set $\hat{\g}=3\frac{\boldsymbol{\na}T}{T}$. In the hydrodynamic phase, one expects that $LT\gg1$ with $\frac{\boldsymbol{\na}T}{T^2}\ll1$, but the change in the matter properties over the matter size is not required to be small $\frac{L\boldsymbol{\na}T}{T}\sim1$. Thus, for relativistic velocities $\frac{|\u|}{1-u_z}\sim1$, our crude estimate indicates that the modification in $\hat{q}$ can be as large as the leading contribution. For instance, taking the same moderate estimates for the transverse flow and temperature gradients as for $L=5\,$fm curve in Fig.~\ref{f:GLVnew}, one readily finds that $\frac{\hat q}{\hat{q}_0}\simeq0.775$.

%
\begin{figure}[t!]
     \centering
     \includegraphics[width=1\textwidth]{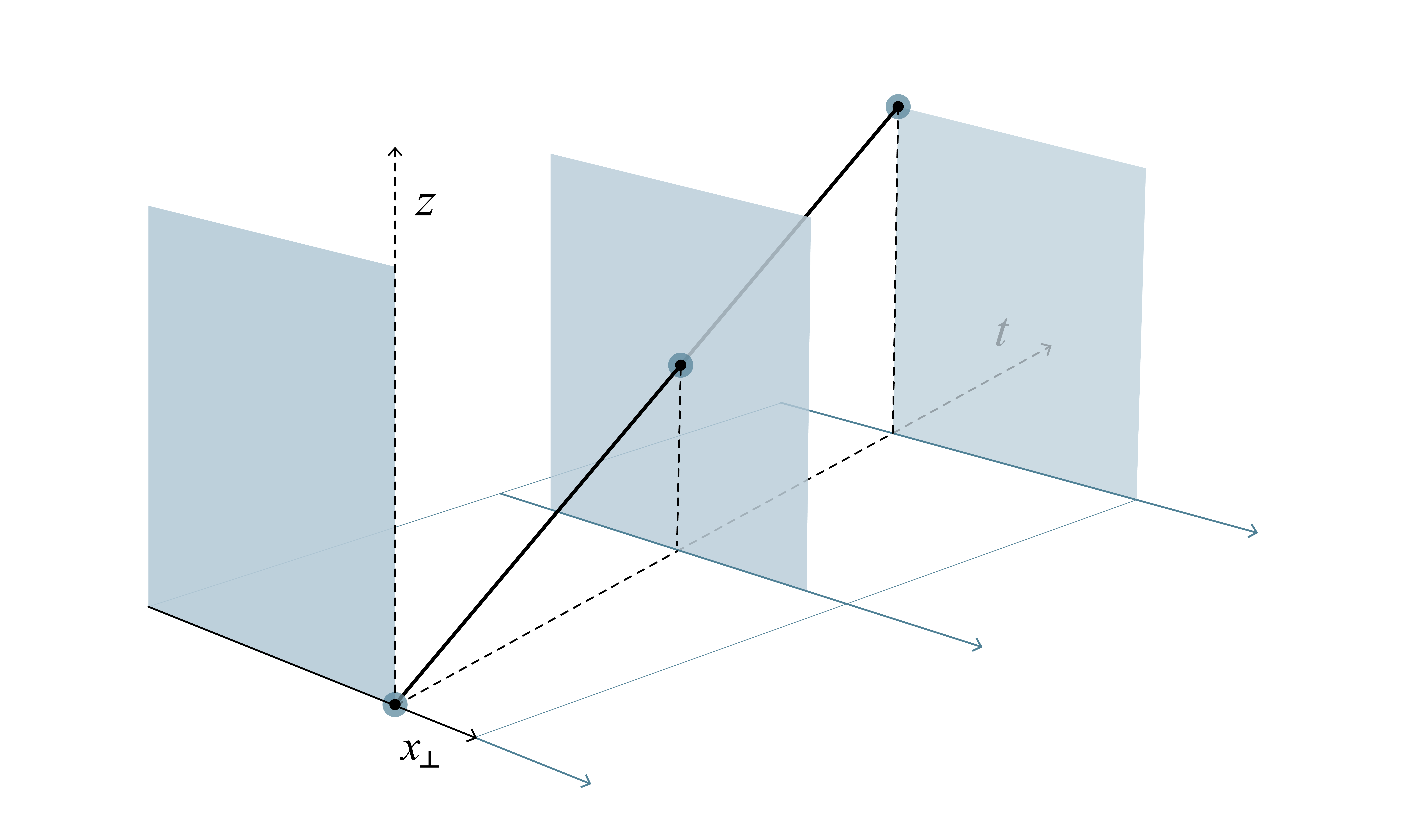}
     \vspace*{-10mm}\caption{The rectangles represent a medium element traveling along the transverse direction, and pictured at three different moments. The leading parton traveling along $z$-direction at $\x=0$ penetrates the matter element at different transverse positions at different times due to the medium transverse motion. Since the medium has transverse structure (introduced by the temperature gradient), the jet sees different local properties, while the matter is assumed to be longitudinally uniform in this illustration.}
\label{f:illustration}
\end{figure}
%

The physical picture behind these larger modifications can be made more transparent if we attempt keeping the full $\x$-dependence in $\rho(\x,z)$, e.g. in \eqref{M_1^2 before expanding}. Assuming that only the Fourier factors are varying fast enough in $\q-\q'$, we find that the corresponding contribution to the amplitude squared is proportional to $\rho\left(-\frac{\u}{1-u_z}z+\frac{\p-\q}{E}z,z\right)$. This change in the local density along the leading parton trajectory agrees with \eqref{e:BroadBorn10} up to first order in gradients, and can be identified with the shift of the matter in the transverse direction over the traveling time $z$, see the illustration in Fig.~\ref{f:illustration}. One should note that the source number density is positive, and higher order gradient corrections ensure that \eqref{e:qhat_modified} is positive. While the other hydrodynamic parameters enter the amplitude squared in a more involved way, the related gradient corrections as well as the full momentum dependence in the integrand are already taken into account in \eqref{dN final} (and in \eqref{dNgu} for the medium-induced branching). 

The presented results should be further included into phenomenological considerations of particular observables. The possible substantial modification in $\hat q$ and the energy loss rate can considerably affect the existing simulations of jets interacting with evolving backgrounds, see e.g. \cite{He:2020iow,Antiporda:2021hpk,He:2015pra,Putschke:2019yrg,Zigic:2021rku,Stojku:2021yow,Barreto:2022ulg,Pablos:2022piv,Singh:2023smw}. It would also be interesting to see if the mixed flow-gradient effects leave clear signatures in more differential observables, discussed in the context of the evolving anisotropic matter, see e.g. \cite{Barata:2023zqg,Andres:2022ovj,Andres:2023xwr,Barata:2023vnl,Andres:2023ymw}. We leave all these considerations for the future work.

\section*{Acknowledgements}
The authors are grateful to P. Arnold, J. Barata, T. Luo, D. Pablos, C. Salgado, and X.N. Wang for multiple discussions and comments on this work. The authors would like to particularly thank M. Sievert and P. Mitkin, who participated in the early discussions shaping this project. The work of XML and AVS is supported by European Research Council project ERC-2018-ADG-835105 YoctoLHC; and by Xunta de Galicia (Centro singular de investigación de Galicia accreditation 2019-2022), by European Union ERDF. XML contribution to this work is supported under scholarship No. PRE2021-097748, funded by MCIN/AEI/10.13039/501100011033 and FSE+. The work of AVS is also supported by the Marie Sklodowska-Curie Individual Fellowship under JetT project (project reference 101032858), and by Funda\symbol{"00E7}\symbol{"00E3}o para a Ci\symbol{"00EA}ncia e a Tecnologia (FCT) under contract 2022.06565.CEECIND. The IGFAE has the CIGUS recognition from the Xunta de Galicia, which accredits the quality and impact of its research. The work of MVK is funded by the Russian Science Foundation Grant 22-22-00664. The work of JR is supported in part by the Mani L. Bhaumik Institute for Theoretical Physics.

\bibliographystyle{bibstyle}
\bibliography{references}

\end{document}